\documentclass[twocolumn]{aa}
\usepackage{epsfig}
\def\simlt{\ \raise -2.truept\hbox{\rlap{\hbox{$\sim$}}\raise5.truept   %
\hbox{$<$}\ }}
\def\simgt{\ \raise -2.truept\hbox{\rlap{\hbox{$\sim$}}\raise5.truept   %
\hbox{$>$}\ }}                                                          %

\def\be{\begin{equation}}
\def\ee{\end{equation}}
\def\newline{\hfil\break}

\def\la{\mathrel{\hbox{\rlap{\hbox{\lower4pt\hbox{$\sim$}}}\hbox{$<$}}}}
\def\ga{\mathrel{\hbox{\rlap{\hbox{\lower4pt\hbox{$\sim$}}}\hbox{$>$}}}}

\def\rxj{RX J1347.5-1145}
\begin{document}
\title{Warming rays in cluster cool cores}
   \author{S. Colafrancesco\inst{1,2} and P. Marchegiani\inst{2,3}}
   \offprints{S. Colafrancesco}
\institute{   ASI-ASDC
              c/o ESRIN, Via G. Galilei snc, I-00040 Frascati, Italy
              Email: colafrancesco@.asdc.asi.it
 \and
              INAF - Osservatorio Astronomico di Roma
              via Frascati 33, I-00040 Monteporzio, Italy.
              Email: cola@mporzio.astro.it
 \and
              Dipartimento di Fisica, Universit\`a di Roma La Sapienza, P.le A. Moro 2, Roma, Italy
              Email: marchegiani@mporzio.astro.it
             }
\date{Received August 7, 2007 / Accepted January 11, 2008 }
\authorrunning {S. Colafrancesco and P. Marchegiani}
\titlerunning {Cosmic rays in cluster cool cores}
\abstract
   {Cosmic rays are confined in the atmospheres of galaxy clusters and, therefore, they can
   play a crucial role in the heating of their cool cores.}
   {
   We discuss here the thermal and non-thermal features of a model of cosmic ray heating of
   clusters' cores that can provide a solution to the cooling-flow problems. To this aim,
   we generalize a model originally proposed by Colafrancesco, Dar \& DeRujula
   (2004) and we show that our model predicts specific correlations between the thermal
   and non-thermal properties of galaxy clusters and enables various observational tests.}
  {
  The model reproduces the observed temperature distribution in clusters by using an
  energy balance condition in which the X-ray energy emitted by clusters is supplied,
  in a quasi-steady state, by the hadronic cosmic rays, which act as ``warming rays'' (WRs).
  The temperature profile of the intracluster (IC) gas is strictly correlated with the pressure
  distribution of the WRs and, consequently, with the non-thermal emission
  (radio, hard X-ray and gamma-ray) induced by the interaction of the WRs with the
  IC gas and the IC magnetic field.}
   {
   The temperature distribution of the IC gas in both cool-core and non cool-core
   clusters is successfully predicted from the measured IC plasma density distribution.
   Under this contraint, the WR model is also able to reproduce the thermal and
   non-thermal pressure distribution in clusters, as well as their radial entropy distribution,
   as shown by the analysis of three clusters studied in details: Perseus, A2199 and Hydra.
   The WR model provides other observable features of galaxy clusters: a correlation
   of the pressure ratio (WRs to thermal IC gas) with the inner cluster
   temperature $(P_{WR}/P_{th}) \sim (kT_{inner})^{-2/3}$, a correlation of the
   gamma-ray luminosity with the inner cluster temperature $L_{\gamma} \sim (kT_{inner})^{4/3}$,
   a substantial number of cool-core clusters observable with the GLAST-LAT experiment,
   a surface brightness of radio halos in cool-core clusters that recovers the observed one,
   a hard X-ray ICS emission from cool-core clusters that is systematically lower than
   the observed limits and yet observable with the next
   generation high-sensitivity and spatial resolution HXR experiments like Simbol-X.}
  {
  The specific theoretical properties and the multi-frequency distribution of the
  e.m. signals predicted in the WR model render it quite different from the other models
  so far proposed for the heating of clusters' cool-cores.
  Such differences make it possible to prove or disprove our model as an explanation of the
  cooling-flow problems on the basis of multi-frequency observations of galaxy clusters.}

 \maketitle

\section{Introduction}
 \label{sec.intro}

Theoretical description of the intra-cluster (IC) plasma in the central regions of many
clusters predicts that it is cooler than in their outskirts, because it radiates X-rays
at such a rate that the plasma cooling time is much shorter than the age of the cluster.
However, such rapid cooling is not observed and the central-temperature depression in
cluster's cores is not as deep as expected on the basis of the plasma's cooling rate: the
observed central temperature $T_{\rm inner}$ settles, in fact, at a fraction $\sim 1/3
-1/2$ of the outer temperature $T_{\rm outer}$ (see, e.g. McNamara 1997,  Peterson et al.
2003, Piffaretti et al. 2005, Donahue et al. 2005, see also Bregman 2004 for a review).
The lack of cooling gas with $T<T_{inner}$ in cluster cores is also supported by the
absence of line emission corresponding to gas below $T_{\rm inner}$, as shown by high
spatial resolution imaging with Chandra (e.g. McNamara et al. 2000; Fabian 2000; Blanton
et al. 2001; Allen et al. 2001; Lewis et al. 2002; Blanton et al. 2003) and high spectral
resolution measurements with the XMM-Newton RGS instrument (Peterson et al. 2001, Tamura
et al. 2001, Kaastra et al. 2001; Kahn et al. 2002; Peterson et al. 2002) and EPIC
instrument (B\"ohringer et al.~2001, 2002; Molendi \& Pizzolato 2001; Matsushita et
al.~2002). These are referred here as the cooling--flow (CF) problems.

It has been widely recognized that some form of heating is necessary to quench the
cooling flow and form a warm--core in the IC gas structure. Available sources of heating
in clusters cores are provided by AGN jets and lobes, pressure waves, buoyant bubbles and
cavities, intra-cluster shock waves, leptonic and hadronic cosmic-rays.\\
The presence of a central radio source with jet and lobes, the mechanical heating
provided by pressure waves produced by the jetted AGNs, the formation of buoyant bubbles
filled with energetic plasma, the thermal conduction from the hot outer layers of the
clusters, may certainly alleviate the CF problems (see e.g., B\"ohringer et al. 2002;
Churazov et al. 2002; Ruszkowski \& Begelman 2002, Fabian 2004, Voit \& Donahue 2005,
Br\"uggen \& Kaiser 2002, Ruszkowski et al. 2004, Vernaleo \& Reynolds 2006, Reynolds et
al. 2005).\\
Some difficulties with solutions along these lines, however, might stand out:
a first difficulty is that not all cooling flow clusters contain a powerful radio
source at their center (e.g., the cluster RXJ0820.9+0752; Bayer-Kim et al.~2002, and the
clusters A1650 and A2244, Donahue et al. 2005), nor a large amount of spherically
distributed buoyant bubbles out to the cluster's core boundary.
Another difficulty is that the bremsstrahlung cooling rate is proportional to $n_{\rm
e}^2\,T^{1/2}$, with $n_{\rm e}$ the plasma's electron number density\footnote{Hydrogen
in the plasma is fully ionized, and we are using here $n_{\rm p}\approx n_{\rm e}$.} and
$T$ its local temperature: a suitable conduction and/or heating and pressure-building
mechanism must somehow adapt itself to this behaviour, particularly as a function of
$n_{\rm e}(r)$, which varies by orders of magnitude along the clusters' radii.
Other difficulties stand out with the spatial distribution of the energy deposited by AGN
jets and cavities: on one hand, the energy released from AGN jets and cavities seems to
be sufficient to quench cooling flow and increase the central gas entropy (e.g., Sijacki
\& Springel 2006); on the other hand, it is not yet clear if such energy can be properly
distributed to reproduce the observed cool-core temperature profiles (see, e.g.,
discussion by Vernaleo \& Reynolds 2005, 2006, Heinz et al. 2006, Vernaleo \& Reynolds
2007). Small duty cicles and a higher fraction of AGNs in cluster cores (see Bird et al.
2007) might however weaken the difficulties related to the AGN heating scenarios. In
addition, there seems to be a small AGN activity in groups (Dwarakanath \& Nath 2006).\\
Sound waves and shocks do not seem to be efficient in quenching cooling flows (see, e.g.,
Fujita \& Suzuki 2006).

Irrespective of the nature of the heating source, it seems that a general property of the
heating agent -- which derives from the smoothness and from the similarity of the heating
distribution required to quench cooling in several clusters -- is that they must be
spatially distributed in the whole cluster core.
In fact, it has been shown (Fabian \& Sanders 2006, Fabian et al. 2006, Sanders \& Fabian
2007) that the non-thermal pressure in the Perseus cluster (one of the best studied
clusters with cool cores and mini radio-halos) is quite similar (slightly steeper) to the
thermal one.\\
In this context, it has been shown that a CR population in clusters that is radially
distributed as the radial profile of the IC gas, is able to recover the observed radial
structure of the IC gas temperature in several clusters (Colafrancesco, Dar \& DeRujula
2004).
The CRs in this last theory are not confined to the disk of the Galaxy, but permeate a
much larger halo, being produced by "cannonballs" travelling for kiloparsecs in the
interstellar medium of galaxies, decelerating by knocking out its constituents, which are
thereby accelerated to become CRs\footnote{When referring to CRs in CF clusters, we use
the expressions ``CRs'' and ``WRs'' as synonyms.} (see, e.g., DeR\'ujula 2004 for a
review, and references therein).\\
Beyond the debate on the cannonball picture (see, e.g., Hillas 2006; Dar \& DeR\'ujula
2006) and its relevance for the CR origin, we stress here that any mechanism which
produces radial distributions of WRs similar to the profile of the IC gas can provide
quenching of cooling flows by the action of WRs heating.

In this paper we consider and justify, on the basis of a consistent picture of CR
diffusion in cluster atmospheres (see Appendix A), that the WR number density $n_{WR}(r)$
can be, in fact, distributed with a radial profile that is similar to that of the IC gas,
and follows the general form $n_{WR} \sim n^{\alpha}_{e}$, i.e. it is proportional to a
power $\alpha$ of the IC gas density $n_e(r)$.
The total amount of WRs in our model and the slope $\alpha$ of the $n_{WR}(r)$ profile
are left as free parameters that will be subsequently constrained by fitting the
temperature profile of the IC gas in each specific cluster core. The resulting CR density
profile will be further tested against the entropy and pressure profiles in the cluster
cores, the spectrum and surface brightness profile of the cluster radio halos and the
future observations in the hard X-ray and gamma-ray bands.\\
We will first show that the presence of WRs in cluster cores are able, in fact, to heat
the IC gas and reproduce their radial temperature profiles.
We will then start from this constraint to discuss the consequences that such a
population of WR hadrons (protons), produced and stored in the atmospheres of galaxy
clusters, has for the gamma-ray emission, for the hard X-ray (HXR) emission and for the
presence of diffuse radio emission of both cool-core and non cool-core clusters.
We will finally show that it is possible to use these predictions, in the framework of a
multi-wavelength observational strategy, to set constraints on the amount and spatial
distribution of WRs in clusters, and therefore disentangle between the WR model discussed
here and other models for the onset of warm-cores in galaxy clusters.\\
The structure of the paper is the following: in Sect.2 we delineate the theoretical
framework that is able to model the temperature structure of the cluster cool cores and
determine the main parameters (i.e., the radial distribution and the density of WRs) that
fit the cluster temperature profiles. In Sect.3 we will also discuss the density and
pressure distribution of WRs in the clusters' atmospheres that is consistent with the
temperature profile.
In Sect.4 we will apply our procedure to a sample of ten well studied clusters. These
clusters have been chosen according to the following criteria: i) 4 clusters which have
cool cores and diffuse radio emission (these clusters certainly have a population of CRs
in their centers); ii) 2 clusters with radio halos (i.e. with CRs in their atmospheres)
but without cool cores, in order to compare their properties with the radio-active,
cool-core clusters; iii) 4 clusters with cool cores but with no evident central diffuse
radio emission in order to compare these non radio--active clusters with the previous
ones.
In Sect.4 we also study the thermal and non-thermal pressure structure of the selected
clusters and we will show that the pressure ratio $P_{WR}/P_{th}$ found at the cluster
center correlates with the central gas temperature $T_{inner}$ providing evidence for the
basic regulation mechanism induced by WRs. We will discuss the predictions of our model
for some specific clusters (like Perseus, A2199 and Hydra) for which detailed thermal and
non-thermal pressure information is available as well as information on their radial
entropy distribution.
In Sect.5 we will discuss the expected gamma-ray luminosity and flux of the cluster
sample here considered that is  correlated with the cool-core temperature structure
predicted in our model.
In Sects.6 and 7 we will present analogous considerations for the expected radio and HXR
emission that are constrained by the cluster temperature structure.
We will discuss the differences between our model and other scenarios for the cool-core
heating in Sect.8 and we will resume our main conclusions in Sect.9.\\
Throughout the paper, we use a flat, vacuum--dominated cosmological model with $\Omega_m
= 0.3$, $\Omega_{\Lambda} = 0.7$ and $h = 0.7$.

\section{Modelling cluster cool cores}
 \label{sec.model}

The virial expectation for the temperature of an un-magnetized plasma in a spherical
cluster with mass $M(r)$ within a radius $r$, is:
\begin{eqnarray}
 k\,T(r)&\simeq& {G\, m_{\rm p}\, M(r)\over 3\, r}\nonumber\\
&\simeq&
          2\;{\rm keV} \left[{M(r)\over 10^{13}\, M_\odot}\right]\,
          \left [{r\over 100\, {\rm kpc}}\right]^{-1}\, ,
\label{Tvirial}
\end{eqnarray}
(see, e.g. Colafrancesco, Dar \& De Rujula 2004).
The observed outer temperature $T_{\rm outer}$ of the IC plasma -- at large radii where
the cooling time should be longer than the cluster's age -- is roughly compatible with
the virial expectation.
In the inner regions of many clusters the temperature is somewhat smaller ($\sim 1/2 -
1/3$ of $T_{\rm outer}$)  than that obtained form Eq.~(\ref{Tvirial}), but significantly
larger than the cooling rate would imply (see, e.g., Peterson et al. 2001).
To reproduce the observed temperature profile from the inner cool region to the outer hot
region, we will consider a population of WRs spatially distributed in the cluster
atmosphere (as obtained by solving a transport equation, see Appendix A) that efficiently
heats the IC gas.

The effect of the ``warming rays'' on a cluster's temperature distribution will be
computed under the following simplifying assumptions:\\
i) spherical symmetry;\\
ii) the radial distribution of the WR density is approximately proportional to a power
$\alpha$ of the thermal plasma density, i.e. $n_{WR} \propto n_{e}^{\alpha}$, with
$\alpha$ and the normalization constant being free parameters. In the Appendix A we show
that the effect of a simple model of cosmic ray diffusion in cluster atmospheres can
provide such kind of radial distribution, for quite general CR source terms;\\
iii) the temperature of the plasma in its ``initial state'' (as the cluster is born) is
approximated by a radius-independent constant value;\\
iv) cooling by X-ray emission and heating and pressure-building by WRs are the dominant
evolutionary agents. Flows and cluster ageing are relatively unimportant.

Given this framework, the rate per unit volume at which the IC plasma loses energy by
thermal bremsstrahlung emission of X-rays writes as:
\begin{eqnarray}
&&{d\epsilon\over dt}\Biggr|_{_{\rm X}}= a\, [n_{\rm e}(r)]^2\, \sqrt{T(r,t)};\nonumber\\
&&a= \sqrt{2^{11}\pi^3\over 3^3}\; {e^6\sqrt{m_{\rm e} }\over h\,m_{\rm e}^2\, c^3}\,
{\bar G}\, {\bar z}\nonumber\\ &&~\sim 4.8\times 10^{-24}\,{\bar z}\,{1\over \sqrt{\rm
keV}} {\rm erg\,cm^3 \over s}\; ,
 \label{Xloss}
\end{eqnarray}
where $\bar z$ is an average charge of the IC plasma (we have approximated here the Gaunt
factor ${\bar G}$ by unity).
We write the IC gas number density as $n_e(r) = n_{e0} \cdot g(r)$, where the function
$g(r)$ contains all the radial dependence of the IC gas density and is obtained from
X-ray observations of each specific cluster.
The rate per unit volume at which the WRs deposit energy in the IC plasma is proportional
to $n_{WR}(r) \cdot n_e(r)$ and it is therefore:
\begin{equation}
{d\epsilon\over dt}\Biggr|_{_{\rm WR}} = b\, \cdot [n_{\rm WR}(r)] \cdot [n_{\rm e}(r)] =
b \cdot n_{\rm WR,0} \cdot n_{\rm e0} \cdot [g(r)]^{1 + \alpha} \; ,
 \label{WRgain}
\end{equation}
since we have assumed that the WR number density writes as $n_{WR}=n_{WR,0} \cdot
[g(r)]^{\alpha}$, i.e. proportional to [$n_{\rm e}(r)]^{\alpha}$.\\
The evolution of the temperature anywhere in the IC plasma satisfies the
energy-conservation relation:
\begin{equation}
3\,k\, n_{\rm e}(r)\, {dT(r,t)\over dt}= {d\epsilon\over dt}\Biggr|_{_{\rm WR}}
 - {d\epsilon\over dt}\Biggr|_{_{\rm X}}\, ,
\label{Tevol}
\end{equation}
with the initial condition $T(r,0)=T_{\rm i}$.
We can rewrite Eq.~(\ref{Tevol}) as:
\begin{equation}
-{dT(r,t)\over  \sqrt{T} - (b/a) [n_{WR}(r)/n_e(r)]}= {a\, n_{\rm e}(r)\over 3\, k}\, dt \label{dT}
\end{equation}
and integrate, to obtain:
\begin{eqnarray}
\ln \left[ {\sqrt{T_{\rm i}/T_{\rm f}}-1\over \sqrt{T(r,t)/T_{\rm f}}-1}\right
]-\sqrt{T(r,t) \over T_{\rm f}}+\sqrt{T_{\rm i} \over T_{\rm f}}=
 \nonumber \\
{b\, n_{\rm WR}(r) n_{\rm e}(r)\over 6 n_{\rm e}(r) k T_{\rm f}}\, (t-t_{\rm i}) \; ,
 \label{Tr2}
\end{eqnarray}
where
 \begin{equation}
kT_{\rm f} \equiv \left[(b/a)(n_{WR,0}/n_{e0})g(r)^{\alpha-1} \right]^2
 \label{eq.tf}
 \end{equation}
is the steady temperature first reached in the central region of the cluster at small
radii $r$ , where the density $n_e(r)$ is high, and at asymptotically late time in the
periphery, where the density $n_{\rm e}(r)$ is vanishingly small. The r.h.s. of
Eq.~(\ref{Tr2}) has been written in a way that shows how the evolution of $T(r,t)$ is
determined by the ratio between the WR deposition rate, $b\, n_{\rm WR}(r) n_{\rm e}(r)$,
and the thermal energy per unit volume, $n_{\rm e}(r) kT_f $, in the inner region where
the cluster's temperature is $T_f$.

\begin{figure}[ht]
\begin{center}
\hspace{-1.cm}
 \epsfig{file=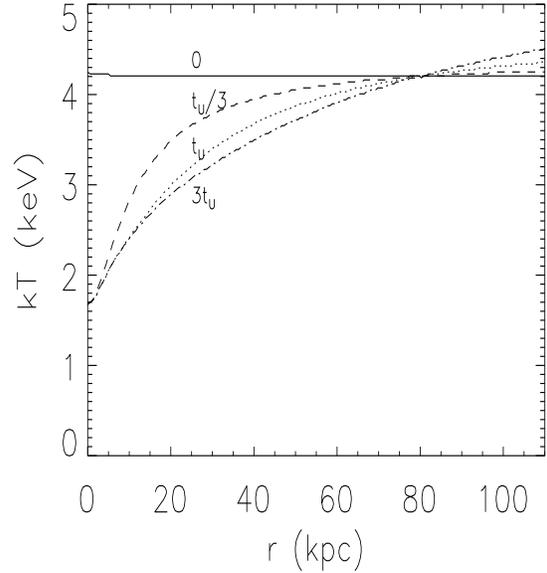,height=9.cm,width=9.cm,angle=0.0}
\end{center}
 \caption{\footnotesize{The evolution of the IC gas temperature profile
 predicted for a cluster similar to A2199 is shown at times $t=0$ (solid), $t_{_{U}}/3$ (dashed),
 $t_{_{U}}$ (dotted) and $3\,t_{_{U}}$ (dot-dashed), with $t_{_{U}}$ being the age of the Universe.
 We assume an initial temperature of $4.2$ keV
 and a density profile equal to that observed in A2199.
 Values $\alpha =0.83$ and $n_{WR,0}=2.3 \cdot 10^{-3}$ cm$^{-3}$ are assumed for this
 specific calculation.}}
 \label{fig.tevo}
\end{figure}

The temperature evolution for a cluster similar to A2199 (we assume here indeed the same
IC gas density profile of A2199 and its outer temperature) as described by
Eq.~(\ref{Tr2}), is shown in Fig.~\ref{fig.tevo} at epoch $t=0,t_{_{U}}/3,t_{_{U}}$ and
$3\,t_{_{U}}$, where $t_{_{U}}$ is the current age of the Universe.
This figure shows that, in our model, a cluster starts as an isothermal system at high
redshifts (i.e., $ t \approx 0$) and then a warm-core develops with time under the
influence of cooling and heating effects balancing each other in order to provide a
stationary configuration at each epoch. We note that the specific choice of the initial
epoch $t_i$ is not very important in the solution of eq.(\ref{Tr2}), because the epoch
$t$ (redshift $z$) at which the cluster is observed is much larger (smaller) than the
initial epoch $t_i$ (redshift $z_i \gg z$). We have nonetheless calculated the
temperature evolution by assuming different values of the initial redshift, $50\leq
z(t_i) \leq 200$, and we found that no appreciable difference appears in the final state
of the temperature radial distribution.

In cool-core clusters the central density is high enough for the equilibrium between
X-ray cooling and WR heating to have been reached at the cluster's center in a time of
the order of its current age. In such a case, $T_{\rm f} \simeq T_{\rm inner}$, i.e. the
presently observed inner-cluster temperature. For all the clusters we study, the cooling
time in the outer parts of the cluster is much longer than $t_{_{U}}$, so that $T_{\rm
i}\simeq T_{\rm outer}$. According to the value of $T_i$, of the IC gas density $n_{\rm
e}$ and of the time $t$, Eq.(\ref{Tr2}) predicts that some clusters have reached this
stage, while others have not.

The procedure here described is used to fit the temperature profile of clusters with and
without cool cores. Given the bremsstrahlung energy loss rate, the temperature radial
profile of each cluster depends on the density and on the spectrum $n_{WR}(E,r)$ of WRs
in clusters.

\section{Warming Rays in cool core clusters}
 \label{sec.wrs}

We consider a spectrum of relativistic WRs (protons) as given by:
\begin{equation}
N_{WR}(E,r)=N_{WR,0} E^{-s} [g(r)]^{\alpha} \; ,
 \label{wr_spectrum}
\end{equation}
with $s$ typically in the range $ \approx 2.3 -3.3$ (see, e.g., Marchegiani et al. 2007),
and we use $s=2.7$ as a reference value, hereafter, for energies larger than $E_{min}=
m_p c^2 \cdot [1+3.4 \cdot 10^{-5}(kT/keV)]$ (see, e.g. Furlanetto \& Loeb 2002). The
value of $E_{max}$ is usually very large and is therefore irrelevant in our calculations
for spectral indices $s
> 2$.

The quantity $N_{WR,0}$ can be derived by assuming that the WR heating rate $b \cdot
n_{WR} \cdot n_e$ provided by the energy lost by the WRs when they interact with the IC
medium, recovers the IC gas temperature profile through eq.(\ref{Tr2}).

The fundamental processes by which the warming rays can heat the IC gas are mainly
Coulomb and hadronic interactions with the plasma nuclei and magneto hydro-dynamical
(MHD) waves excited by CR proton anisotropies.

For moderately relativistic CR nuclei with a charge $z$, velocity $v\!=\!\beta\, c$ and
Lorentz factor $\gamma$, the Coulomb energy-loss rate is:
\begin{equation}
-{dE\over dt}\approx K \, z^2\, Z^2\,{1\over \beta}\,
   \left [\ln {2\, m_{\rm e}\, c^2\, \beta^2 \gamma^2\over I_{\rm p}}
 -\beta^2\right],
\label{dedx}
\end{equation}
where $Z^2$ is the (suitably averaged) squared charge of the plasma's nuclei, $K\!=\!4\,
\pi\, n_{\rm e}\, r_{\rm e}^2\, m_{\rm e}\, c^3$ and $r_{\rm e}\!=\!e^2/m_{\rm e}\,
c^2\!\simeq\!2.82$ fm.
Here $I_{\rm p}\!=\!\hbar\, \omega_{\rm p}$, with $\omega_{\rm p}\!=\! [4\pi\,
n_{\rm e}\,e^2/m_{\rm e}]^{1/2}$ the plasma frequency.%
\footnote{It has been noticed by Colafrancesco, Dar \& DeRujula (2004) that a highly
ionized plasma is much more efficient than non-ionized gas in slowing down the CR nuclei.
Energy losses via Coulomb collisions with neutral atoms become significant only for
energy transfers larger than their ionization potential $I$. The corresponding small
impact parameters imply relatively small cross sections. For the IC plasma, the role of
$I$ is played by the much smaller quantity $I_{\rm p}$.}
The heating rate given in Eq.~(\ref{dedx}), for typical plasma densities near the center
of clusters ($n_{\rm e} \!\sim\!10^{-2}$ cm$^{-3}$), is $\sim\! 4$ times larger than for
neutral hydrogen.
For protons ($z\!=\!1$) with a typical WR energy $(\gamma\!=\!2)$ and for a plasma
consisting of $\sim\! 0.93\%$ H, $\sim\! 0.07\%$ He (by number) and traces of heavier
elements, one finds $Z^2\!\sim\! 1.14$ and $dE/dt\approx 10^{-17}$ GeV/s, for the quoted
electron number density.

Hadronic collisions between WRs and plasma nuclei constitute an energy-loss mechanism
 which is approximately as efficient as that given in Eq.~(\ref{dedx}). Here we consider
the collisions of WR protons with ambient protons, the discussion of proton--nucleon and
nucleon--nucleon collisions being far too cumbersome to be justified by the small
admixtures of nuclei heavier than protons in the WR and IC constituencies. For values
$\gamma>2$ of the incident proton, the $pp$ cross section is $\sigma_{pp}\!\sim\! 4\cdot
10^{-26}$ cm$^2$, and it is dominantly inelastic. In practically all of these collisions,
the incident proton survives unscathed, as the ``leading'' (most energetic) final-state
particle, carrying on average $\approx 70\%$ of the incident proton's energy. Thus, a
proton's energy-loss rate by hadronic collisions in a hydrogenic plasma is
\begin{equation}
-{dE\over dt}\sim 0.7\,n_{\rm e}\,\sigma_{pp}\,c\;E \;
 \label{dedx2}
\end{equation}
(Mannheim \& Schlickeiser 1994), that reads $dE/dt\!\sim\!1.6\times 10^{-17}$ GeV/s, for
our reference value of $n_{\rm e}$ and $\gamma\!=\!2$. This is very close to the Coulomb
energy loss previously computed for the same reference parameters.

An additional process which has been discussed in the literature is the effect of the MHD
waves excited by proton anisotropy, which might occur as relativistic, charged particle
stream faster than the local Alfven speed (see, e.g. Skilling 1971); these waves can be
dissipated efficiently in the background ICM via Landau damping and other dissipative
mechanisms (see, e.g., Foote \& Kulsrud 1979). The effect of this mechanism in cluster
atmospheres has been studied in details by Loewenstein et al. (1991): under the
assumptions that the wave velocity is large enough to heat the IC gas significantly, and
that the Coulomb and hadronic heating can be neglected, these authors found that the
thermal pressure becomes overwhelmed by the CR proton pressure. Therefore, they concluded
that such CR heating mechanism alone is not able to stabilize the cooling flows.

Following the previous considerations, we assume therefore that the Coulomb and hadronic
heating are the dominant sources and we search for the amount and spatial distribution of
WRs required to produce the temperature structure observed in a sample of clusters with
cool cores (i.e. where heating is required) and radio halos (i.e., where CRs are
required). In Sect. \ref{par.pressioni} we will check our results by comparing the
non-thermal and thermal pressures predicted in our model with those observed in the same
clusters.

The total WR loss rate is given by the sum of eqs. (\ref{dedx}) and (\ref{dedx2})
\begin{equation}
\bigg(\frac{dE}{dt}\bigg)_{tot} \simeq n_e (A_{coul} +A_{hadr})
\end{equation}
with
\begin{eqnarray}
 A_{coul} & = & 4 \pi r_e^2 m_e c^3 z^2 Z^2 {1 \over \beta} \left [\ln {2\, m_{\rm e}\, c^2\, \beta^2 \Gamma^2\over I_{\rm p}}
 -\beta^2\right] \nonumber \\
 A_{hadr} & \simeq & 0.7\,\sigma_{pp}\,c\;E \; ,
 \label{eq.a}
\end{eqnarray}
and is approximately proportional to $n_e$ (by neglecting the logarithmic dependence on
$n_e$ contained in $\omega_p$ in eq.\ref{dedx}).

The value of $b$ (see eq.\ref{WRgain}) is found by convolving the energy loss rate with
the WR spectrum and equating with the warming rate
\begin{equation}
 {d \epsilon \over dt} \Biggr|_{_{\rm WR}} \equiv \int_{E_{min}}^{E_{max}} N_{WR}(E,r)
 \bigg({dE \over dt}\bigg)_{tot} dE \; ,
\end{equation}
which writes as
\begin{equation}
b \cdot n_{\rm WR}(r) \cdot n_e(r)
  \equiv N_{WR,0} g^{\alpha}(r) n_e(r) I_A \; ,
\end{equation}
where
\begin{equation}
I_A \equiv \int_{E_{min}}^{E_{max}} E^{-s} (A_{coul}+A_{hadr}) dE \; .
 \label{eq.Ia}
\end{equation}
Since the WR density at the cluster's center is given by
 \be
 n_{\rm WR,0} = \int_{E_{min}}^{E_{max}} dE N_{WR,0} E^{-s} \; ,
 \label{eq.nwr0}
 \ee
we obtain
\begin{equation}
 b= {I_A \over \int_{E_{min}}^{E_{max}} dE E^{-s}}.
\label{eq.b}
\end{equation}
Inserting the previous expression of $b$ in eq.(\ref{Tr2}) we obtain the value of $n_{\rm
WR,0}$ required to fit the temperature profile, and in turn the pressure of the WRs as
\begin{equation}
 P_{WR}(r)=\frac{1}{3}m_pc \int_{\gamma_{min}}^{\gamma_{max}} N_{\rm WR}(\gamma,r) \beta
 \gamma v(\gamma) d\gamma \; ,
 \label{eq.press.cr}
\end{equation}
where $\gamma=E_p/(m_pc^2)$.

\section{The effect of WRs in cluster cores}

\subsection{Temperature profiles}

The solution of eq.(\ref{Tr2}) provides a fit to the cluster's temperature profile by
adjusting the free parameters $\alpha$ and $n_{WR,0}$. We have applied this analysis to
ten well studied clusters, eight of which are cool-core (CC) clusters and two are non
cool-core (NCC) clusters (see Table \ref{tab.1}).
Specifically, we considered 4 clusters with cool-cores which have central radio emission
(usually of the class of mini radio-halos): these are A2199, Perseus, A2390 and \rxj.
For comparison, we also considered two well known non cool-core clusters with radio halos
(Coma and A2163) and some of the best studied clusters with cool-cores but no evident
diffuse radio emission, i.e. A262, A133 (which has a relic radio emission feature), Hydra
and A1795.
\begin{table*}[htb]{}
\vspace{2cm}
\begin{center}
\begin{tabular}{|*{6}{c|}}
\hline
 Cluster & $z$ & $T_{\rm inner}$ & $T_{\rm outer}$  & Notes & Ref.\\
         &     &  keV            &  keV             &       &     \\
         \hline
 A262 & 0.0162 & 0.95 & 2.4 & CC & [1] \\
 A2199 & 0.030 & 1.6 & 4.2  & CC; RH & [2] \\
 A133 & 0.057 & 2.05 & 4.61 & CC; Relic & [1] \\
 Perseus & 0.0179 & 3.04 & 6.7 & CC; RH & [3] \\
 Hydra & 0.0539 & 3.1 &  4.0 & CC & [4] \\
 A1795 & 0.0625 & 3.5 &  6.5 & CC & [1] \\
 A2390 & 0.2304 & 4.89 & 9.81 & CC; RH & [1] \\
 \rxj & 0.451 &  7.04 & 19.4 & CC; RH & [5] \\
\hline
 Coma & 0.023 & 8.2 & 8.2 & NCC; RH & [6] \\
 A2163 & 0.203 & 14.6 & 14.6 & NCC; RH & [7] \\
 \hline
 \end{tabular}
 \end{center}
 \caption{\footnotesize{The list of clusters analyzed.
 Redshift $z$, inner observed temperature $T_{\rm inner}$ and outer observed temperature
$T_{\rm outer}$ are given. The last two columns show which cluster has a radio halo (CC=
cool-core; NCC= non cool-core; RH= radio halo) and the references of the data we used for
the fit. [1] Vikhlinin et al. (2005); [2] Johnstone et al. (2002); [3] Churazov et al.
(2003); [4] David et al. (2001); [5] Allen et al. (2002); [6] Briel et al. (1992); [7]
Elbaz et al. (1995).
 }}
 \label{tab.1}
 \end{table*}

For each cluster in Tab.\ref{tab.1}, we found the values of $\alpha$ and $n_{\rm WR,0}$
which best fit the observed temperature profile, by assuming that initially (i.e. at high
$z$ when the cluster formed) each cluster was isothermal with temperature $T_i$ equal to
the outer observed temperature, and that the evolution of the temperature proceeds
according to eq.(\ref{Tr2}) and finishes at the redshift $z$ at which each cluster is
observed.\\
The WR density $n_{\rm WR}(r)$ found through this procedure is the one required to heat
the IC gas at the level shown by the temperature profile by using eq.(\ref{Tr2}).

We notice that the WR density in cool-core clusters is required to be flatter (i.e. with
values $\alpha < 1$) than the IC gas density to provide the heating rate necessary to
quench the gas cooling and reproduce the cluster's temperature profile. On the other
hand, clusters without cool-cores require $\alpha \approx 1$ because they did not yet
developed gas cooling conditions (see discussion at the end of Sect.2).

\subsection{Pressure and density of WRs}

The ratio of the central WR pressure $P_{WR}$ to the IC gas thermal pressure $P_{th}$ is
found by using eq.(\ref{eq.press.cr}) and the expression $P_{th}= 2 n_e(r) kT(r)$ (by
including the contribution of both thermal protons and electrons). The value of the
best-fit central pressure ratio $P_{WR}/P_{th}$ found for each cluster considered in this
study is reported in Table \ref{tab.2}.
\begin{table*}[htb]{}
\vspace{2cm}
\begin{center}
\begin{tabular}{|*{7}{c|}}
\hline
 Cluster & $\alpha$ & $n_{\rm WR,0}$    & $P_{WR}/P_{th}$ & $F_\gamma$         & $L_\gamma$ & $F_{HXR}$  \\
         &          & cm$^{-3}$ &                        & cm$^{-2}$ s$^{-1}$ & erg s$^{-1}$ & erg cm$^{-2}$ s$^{-1}$\\
         \hline
 A262 & 0.83 & $2.20\times 10^{-3}$ & 1.23 & $3.89\times 10^{-9}$ & $1.43\times10^{42}$ & $3.87\times10^{-14}$ \\
 A2199 & 0.83 & $2.31\times 10^{-3}$ & 0.92 & $8.43\times 10^{-9}$ & $1.08\times10^{43}$ & $3.06\times10^{-13}$ \\
 A133 & 0.84 & $4.56\times 10^{-4}$ & 0.77 & $7.30\times 10^{-10}$ & $3.53\times10^{42}$  & $6.10\times10^{-15}$\\
 Perseus & 0.91 & $4.98\times 10^{-4}$ & 0.54 & $2.20\times 10^{-8}$ & $9.91\times10^{42}$ & $1.59\times10^{-13}$\\
 Hydra & 0.97 & $6.24\times 10^{-4}$ & 0.57 & $3.46\times 10^{-9}$ & $1.49\times10^{43}$ & $2.57\times10^{-14}$\\
 A1795 & 0.96 & $5.55\times 10^{-4}$ & 0.50 & $3.17\times 10^{-9}$ & $1.86\times10^{43}$ & $2.41\times10^{-14}$\\
 A2390 & 0.94 & $2.21\times 10^{-4}$ & 0.41 & $1.41\times 10^{-10}$ & $1.39\times10^{43}$ & $6.17\times10^{-16}$\\
 \rxj & 0.89 & $9.20\times 10^{-4}$ & 0.33 & $5.07\times 10^{-10}$ & $2.37\times10^{44}$ & $8.66\times10^{-16}$\\
 Coma & 1.00 & $1.07\times 10^{-5}$ & 0.29 & $1.65\times 10^{-8}$ & $1.23\times10^{43}$ & $1.67\times10^{-13}$\\
 A2163 & 1.00 & $1.42\times 10^{-5}$ & 0.21 & $1.31\times 10^{-9}$ & $9.71\times10^{43}$ & $4.85\times10^{-15}$\\
 \hline
 \end{tabular}
 \end{center}
 \caption{\footnotesize{
 Col.1: cluster name;
 Col.2: $\alpha$;
 Col.3: $n_{\rm WR,0}$  (in units of cm$^{-3}$);
 Col.4: the pressure ratio $P_{WR}/P_{th}$ at the cluster center;
 Col.5: the gamma-ray flux $F_{\gamma}$ in units of cm$^{-2}$ s$^{-1}$;
 Col.6: the gamma-ray luminosity $L_{\gamma}$ in units of erg s$^{-1}$.
 (The gamma-ray luminosity and flux are integrated in the 0.1--10 GeV band).
 Col.7: the HXR flux $F_{HXR}$ in units of erg cm$^{-2}$ s$^{-1}$ integrated
 in the 10--50 keV band.
 }}
 \label{tab.2}
 \end{table*}

\begin{figure}[ht]
\begin{center}
 \epsfig{file=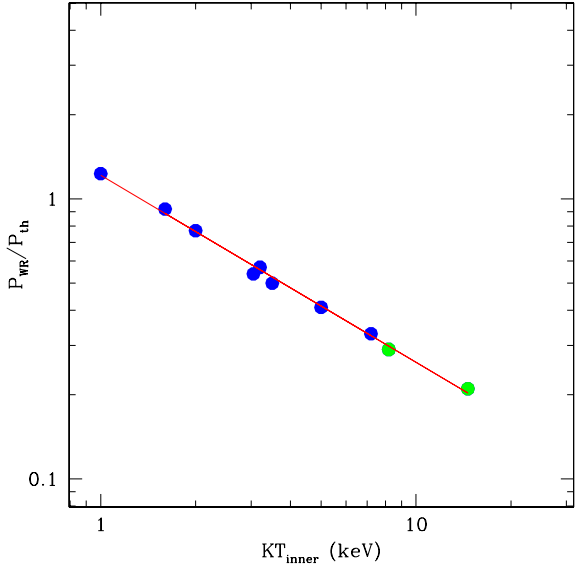,height=8.cm,width=8.cm,angle=0.0}
\end{center}
 \caption{\footnotesize{The correlation of the central pressure ratio $P_{WR}/P_{th}$ with the
 inner temperature of the cool core $T_{inner}$ in keV is given for each cluster in Tab.1.
 The fit $P_{WR}/P_{th} \propto (k T_{inner})^{-2/3}$ is shown as a solid line.
 Blue dots refer to cool-core clusters and green dots to non cool-core clusters.}}
 \label{fig.p_t1}
\end{figure}

\begin{figure}[ht]
\begin{center}
\hspace{-1cm}
 \epsfig{file=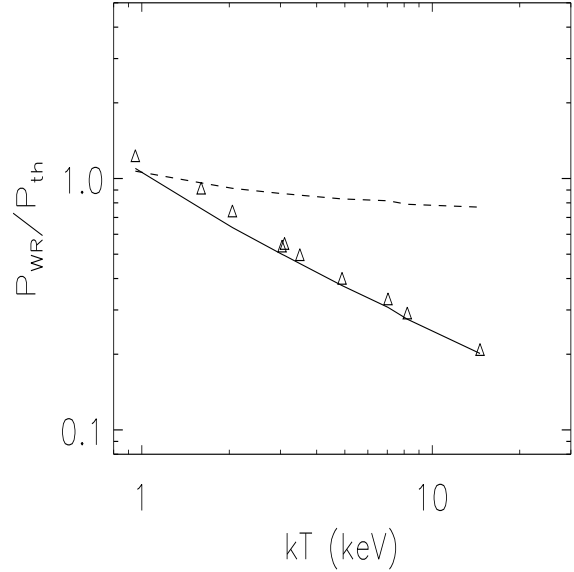,height=9.cm,width=9.cm,angle=0.0}
\end{center}
 \caption{\footnotesize{The contribution of the term $({a
\over b}) \langle \varepsilon_{WR} \rangle$ (dashed curve) to the correlation of the
pressure ratio $P_{WR}/P_{th}$ with the inner temperature of the cool core $T_{inner}$ is
shown. The solid curve is the convolution of the terms $({a \over b}) \langle
\varepsilon_{WR} \rangle$ and $(kT_{inner})^{-1/2}$ (see text for details).
 }}
 \label{pratio_scaling}
\end{figure}

The results of our fitting procedure to the cluster temperature profiles indicate that
there is a strong correlation between the inner cluster temperature $T_{inner}$ and the
pressure ratio $P_{WR}/P_{th}$ at the cluster center (see Fig. \ref{fig.p_t1}). This
correlation can be well represented with a power-law function $P_{WR}/P_{th} \propto (k
T_{inner})^\delta$ with $\delta \approx - 2/3$.\\
Such correlation is due to the fact that in clusters with cooler inner temperature, a
high IC gas density triggers fast gas cooling; therefore, a stronger warming action
(compared to the thermal energy) is required to quench the cooling and to adjust the
temperature profile to the observed one by the energy-conservation relation in
eq.(\ref{Tr2}).
The central pressure ratio $P_{WR}/P_{th}$ that is required in order to equilibrate gas
cooling and then settle the IC gas at the inner temperature value $T_{inner}$ is obtained
from eq.(\ref{eq.tf}) and reads
\begin{eqnarray}
 P_{WR}/P_{th} & =& {1 \over 2 \sqrt{kT_{inner}}} \bigg( { a \over b} \bigg) {P_{WR} \over n_{WR,0}}
 \nonumber \\
  & = &  {1 \over 2 \sqrt{kT_{inner}}} \bigg( { a \over b} \bigg) \langle
  \varepsilon_{WR}\rangle
  \label{eq.pratio}
\end{eqnarray}
The correlation $P_{WR}/P_{th} \propto (k T_{inner})^{-2/3}$ is explained by the combined
temperature dependence of the term $\propto (kT_{inner})^{-1/2}$ and of the quantity $({a
\over b}) \langle \varepsilon_{WR} \rangle \! \sim \! (kT_{inner})^{-1/6}$, which depends
on $T_{inner}$ through the lower limit of integration $E_{min}(T_{inner})$ in the
expressions of $b$ and $ \langle\varepsilon_{WR} \rangle$ (see Fig.\ref{pratio_scaling}).
Note that the $T_{inner}$ dependence of $b$ and $\langle \varepsilon_{WR} \rangle$ arises
only from the spectral properties of WRs in our model, while the scaling with
$(kT_{inner})^{-1/2}$ arises from the condition set in eq.(\ref{Tr2}).

\begin{figure}[ht]
\begin{center}
 \epsfig{file=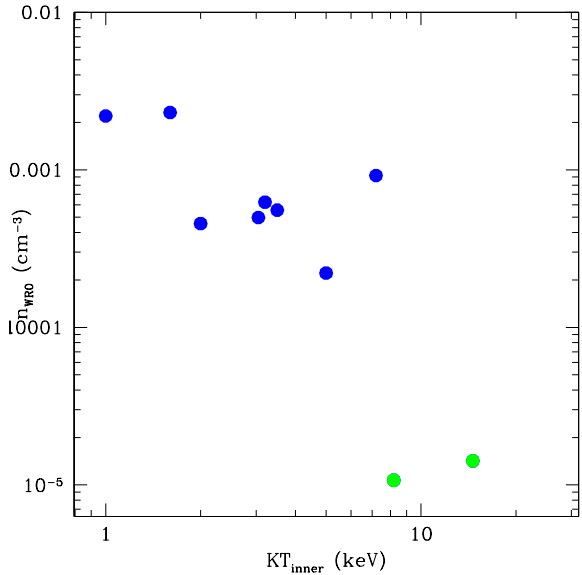,height=8.cm,width=8.cm,angle=0.0}
\end{center}
 \caption{\footnotesize{The correlation of the central WR density $n_{\rm WR,0}$
 with the inner temperature of the cool core $T_{inner}$.
 Blue dots refer to cool-core clusters and green dots to non cool-core clusters.}}
 \label{fig.nwr0_t}
\end{figure}

\begin{figure}[ht]
\begin{center}
\hspace{-1cm}
 \epsfig{file=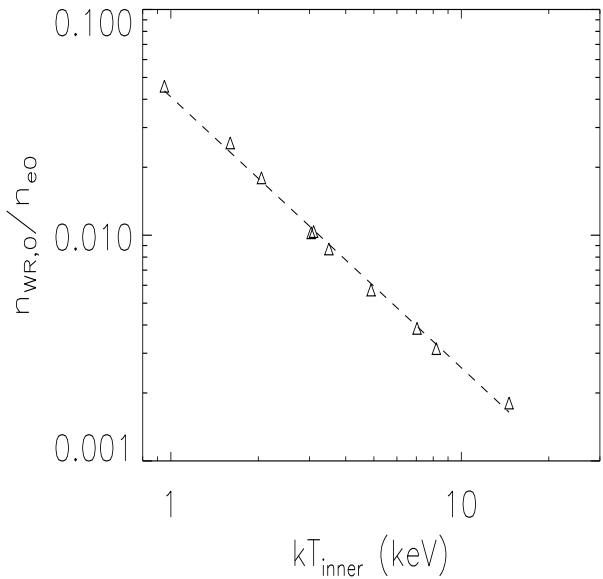,height=9.cm,width=9.cm,angle=0.0}
\end{center}
 \caption{\footnotesize{The correlation of the ratio $n_{\rm WR,0}/n_{\rm e0}$
 with the inner temperature of the cool core $T_{inner}$.
 The dashed line shows the trend $n_{WR,0}/n_{e0} \propto (kT_{inner})^{-1.2}$.}}
 \label{fig.densrapp_t}
\end{figure}

The other correlation, which is naturally related to the previous one, is found -- as
expected -- between the central WR density $n_{\rm WR,0}$  and the value of $k T_{inner}$
(see Fig.\ref{fig.nwr0_t}).
In fact, from eq.(\ref{eq.tf}) we obtain that
\begin{equation}
kT_f=\left ( \frac{b}{a} \cdot \frac{n_{WR,0}}{n_{e0}} \right)^2.
\end{equation}
for $r\rightarrow 0$.
Since the WRs are normalized to obtain $T_f \sim T_{inner}$, we can write
\begin{equation}
n_{WR,0} \sim a \sqrt{kT_{inner}} \cdot n_{e0} \cdot \frac{I_E}{I_A}
\end{equation}
where $I_E= \int_{E_{min}}^{E_{max}} dE E^{-s}$ and $I_A$ is defined in eq.(\ref{eq.Ia}).
Then the ratio $I_E/I_A$ depends on $E_{min}$ and, as a consequence, on $kT_{inner}$; we
found that for our range of temperatures this dependence is $I_E/I_A \propto
(kT_{inner})^{-1.7}$. The conclusion is that
\begin{equation}
\frac{n_{WR,0}}{n_{e0}} \propto (kT_{inner})^{-1.2},
\end{equation}
which is confirmed by our results (see Fig. \ref{fig.densrapp_t}).

\subsection{Pressure profiles: the cases of Perseus and A2199}
\label{par.pressioni}

We compare the pressure profiles of the (non-thermal) WRs and of the thermal components
obtained in our model by fitting the IC gas temperature profile to the data available for
Perseus and A2199.

\begin{figure}[ht]
\begin{center}
\hspace{-1cm}
 \epsfig{file=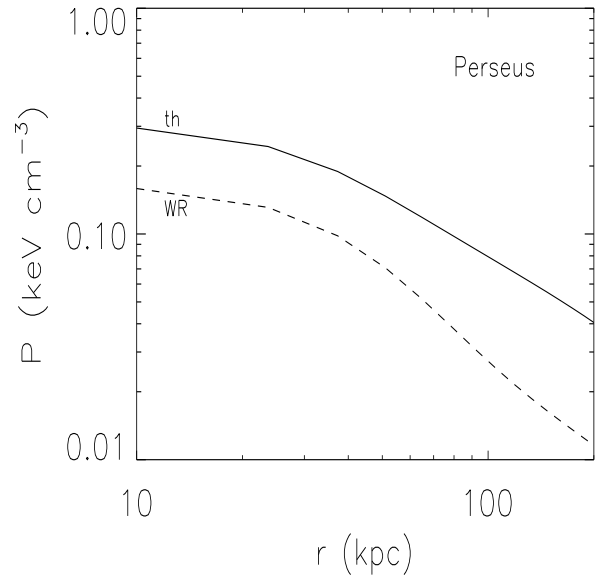,height=9.cm,width=9.cm,angle=0.0}
\end{center}
 \caption{\footnotesize{The radial profile of the WR pressure (dashed curve)
 of Perseus is compared
 with that of the thermal gas (solid curve). The radial profile of the WR pressure resembles the radial
 profile of the non-thermal pressure found by Sanders \& Fabian (2007).}}
 \label{fig.perseus_p}
\end{figure}

For Perseus we find that a central ratio $P_{WR}/P_{th} \approx 0.54$ is required in
order to reproduce the inner temperature profile; such a value is very well consistent
with that found between the non-thermal and the thermal pressures in the analysis of
Sanders \& Fabian (2007). It is also remarkable that our model is able to reproduce the
steeper spatial profile of the non-thermal pressure w.r.t. the thermal pressure profile
found by these authors in their analysis of the central regions of Perseus (see
Fig.\ref{fig.perseus_p}). We note also that the non-thermal pressure we find in this
cluster is lower than the thermal one: in this respect, we can conclude that our model
has not the problems shown by scenario of CR heating provided by waves excited by protons
anisotropy (see Loewenstein et al. 1991).
In our framework, the non-thermal pressure component found in the core of Perseus can be
entirely attributed to WRs.
These results bring our WR heating model to a strong consistency level with the results
of the available X-ray observations (e.g. Sanders \& Fabian 2007).
As a consequence of the presence of WRs in the core of Perseus cluster, we expect a
non-thermal HXR emission due to ICS of the secondary electrons from the central region of
Perseus: we will discuss in Sect.7 the predictions of our model for the HXR emission in
the 2-10 keV band where the instruments on board Chandra and XMM-Netwon are sensitive.

We have also studied the energy density profile of the WRs and of the thermal IC gas in
A2199. The WR energy density has been evaluated as
 \be
 E_{WR}(r) = m_p c^2 \int_{\gamma_{min}}^{\gamma_{max}} N_{\rm WR}(\gamma,r)
 (\gamma -1) d\gamma
 \label{eq.energy.cr}
 \ee
and again depends on the WR spatial distribution that fits the cluster temperature
profile. The thermal IC gas energy density is evaluated as $E_{th} = 3 n_e(r) kT(r)$ (by
including the contribution of both thermal protons and electrons).
\begin{figure}[ht]
\begin{center}
\hspace{-1cm}
 \vbox{
 \epsfig{file=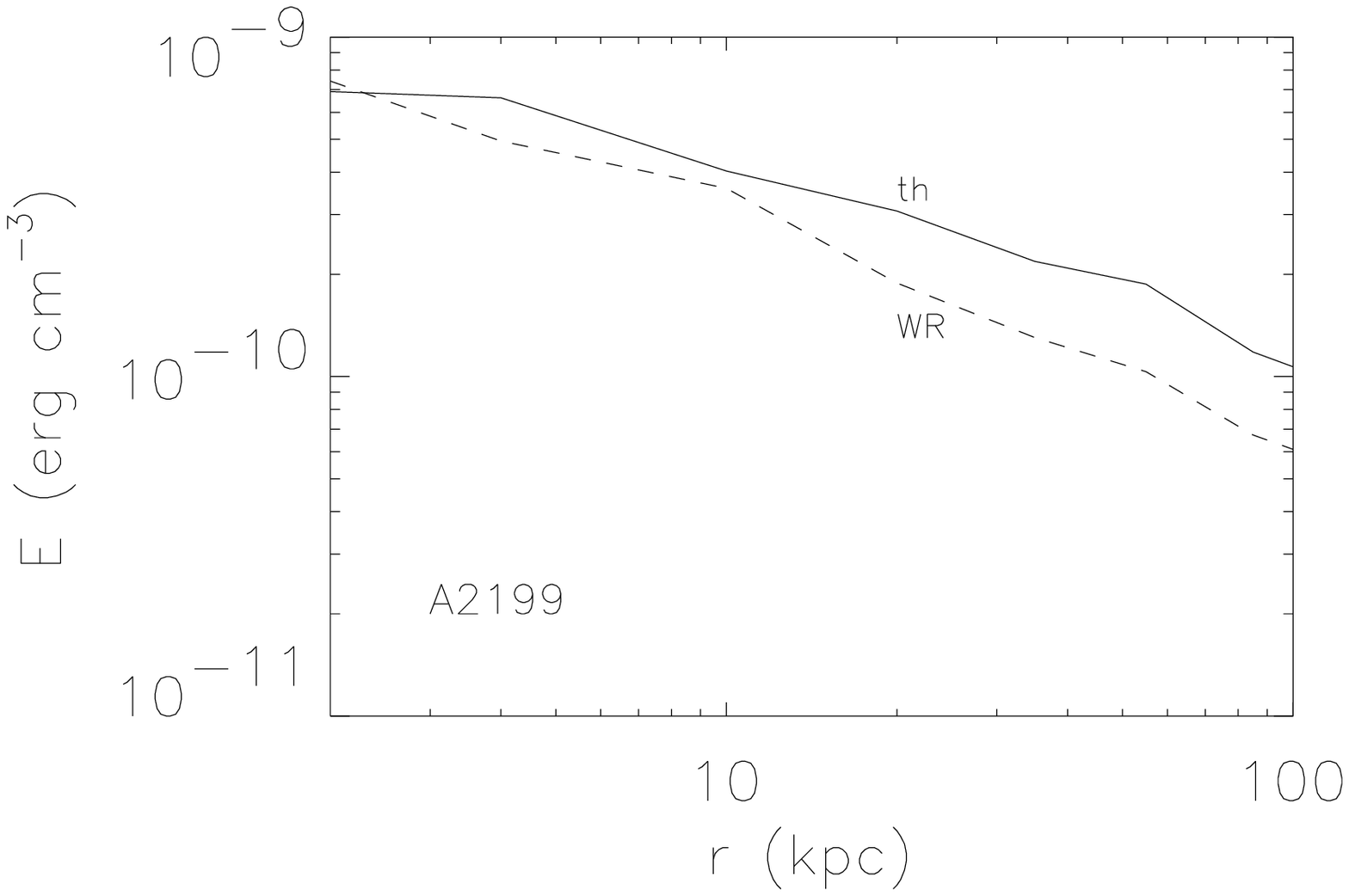,height=9.cm,width=9.cm,angle=0.0}
 \epsfig{file=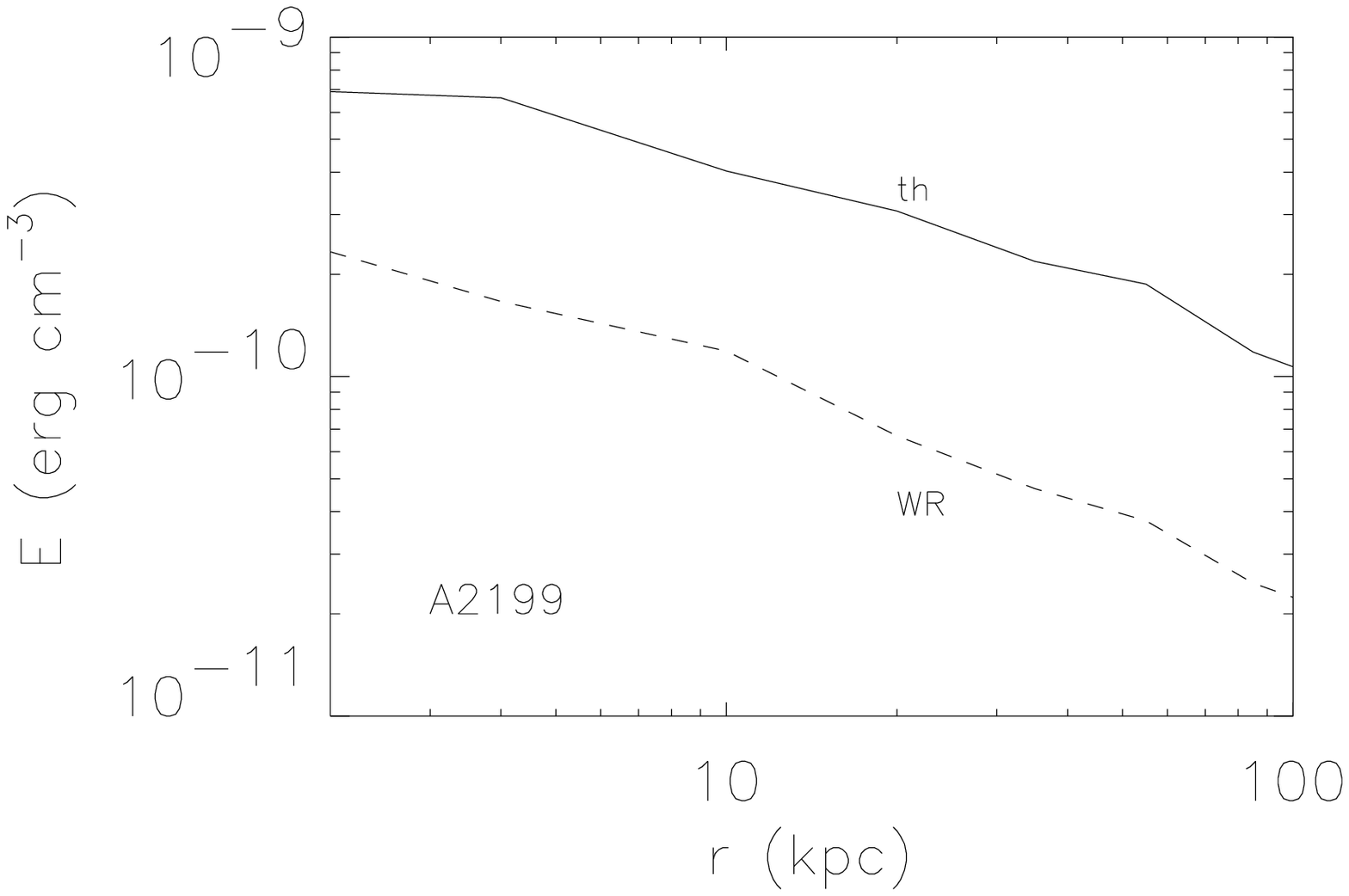,height=9.cm,width=9.cm,angle=0.0}
 }
\end{center}
 \caption{\footnotesize{The radial profile of the WR energy density of A2199 is compared
 with that of the thermal gas. The two panels refer to i) a proton spectrum
with $\gamma_{min}=1+3.4\times10^{-5} (kT/keV)$ (see, e.g., Furlanetto \& Loeb 2002); ii)
a proton spectrum with $E_{kin,min}=706$ MeV.}}
 \label{fig.A2199_p}
\end{figure}
The comparison of the two energy densities as evaluated in our model is shown in the
upper panel of Fig.\ref{fig.A2199_p}. Our model predicts that the WR energy density at
the center of A2199 is approximately equal to the IC gas thermal energy in order to fit
the cluster's temperature profile. Similarly to the case of Perseus, we find a radial
profile of the WR energy density that decays faster with radius with respect to the
thermal energy density.\\
It is possible to recover the energy density ratio of A2199 as presented in the model of
Guo \& Oh (2007) by noticing that the minimum energy cutoff chosen by these last authors
is $\approx 706$ MeV, i.e. much higher than the value of $E_{min}$ assumed in this paper
(see Sect.3). In fact, when we cut the WR spectrum at this higher value of $E_{min}$, we
recover an energy density ratio which is quite close to the value found in Guo \& Oh
(2007). We verified that the residual difference in this ratio is due to the marginal
effect of the thermal conduction in their model.

\subsection{Entropy profiles}

\begin{figure}[ht]
\begin{center}
\hspace{-1cm}
 \vbox{
 \epsfig{file=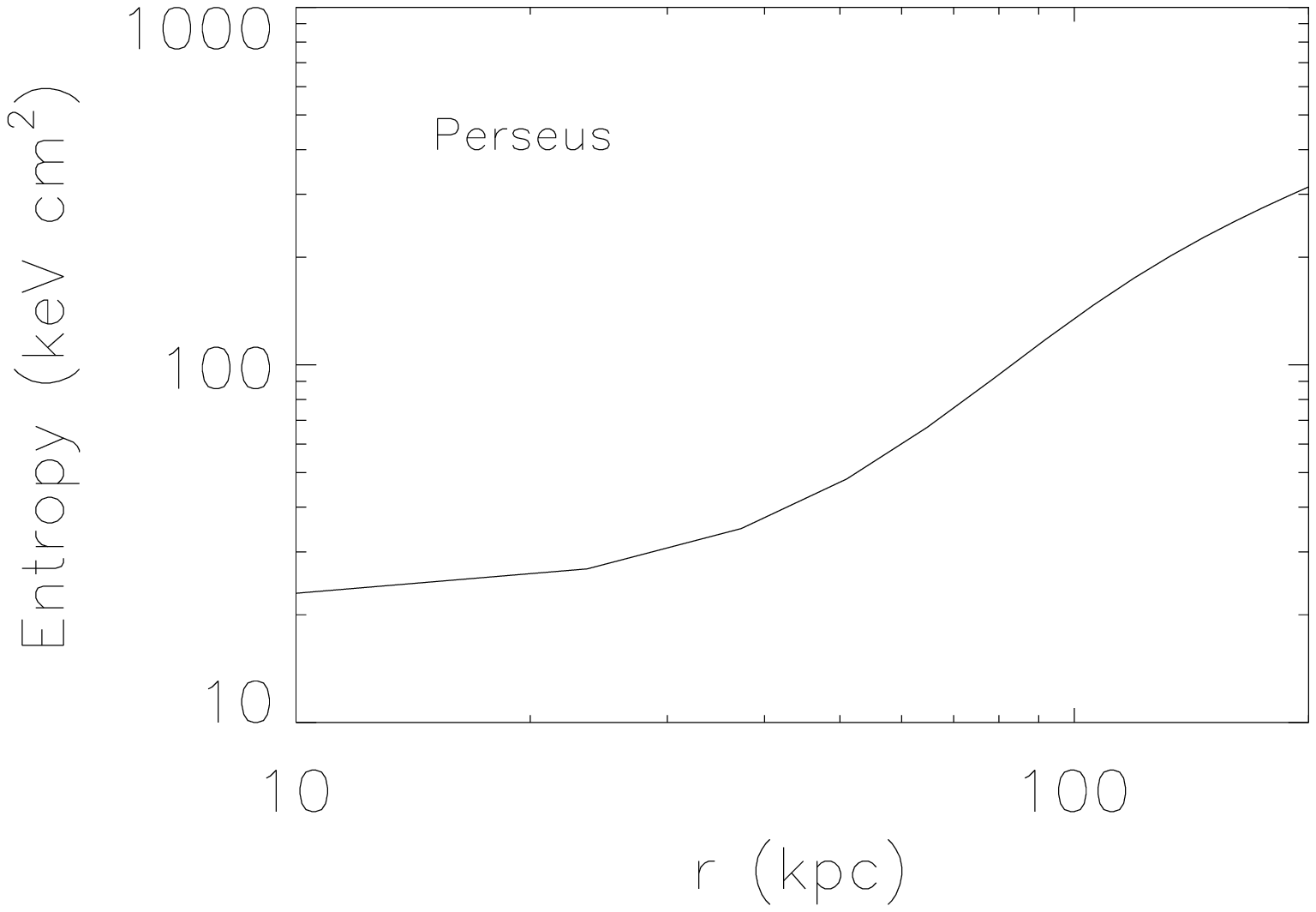,height=5.5cm,width=9.cm,angle=0.0}
 \epsfig{file=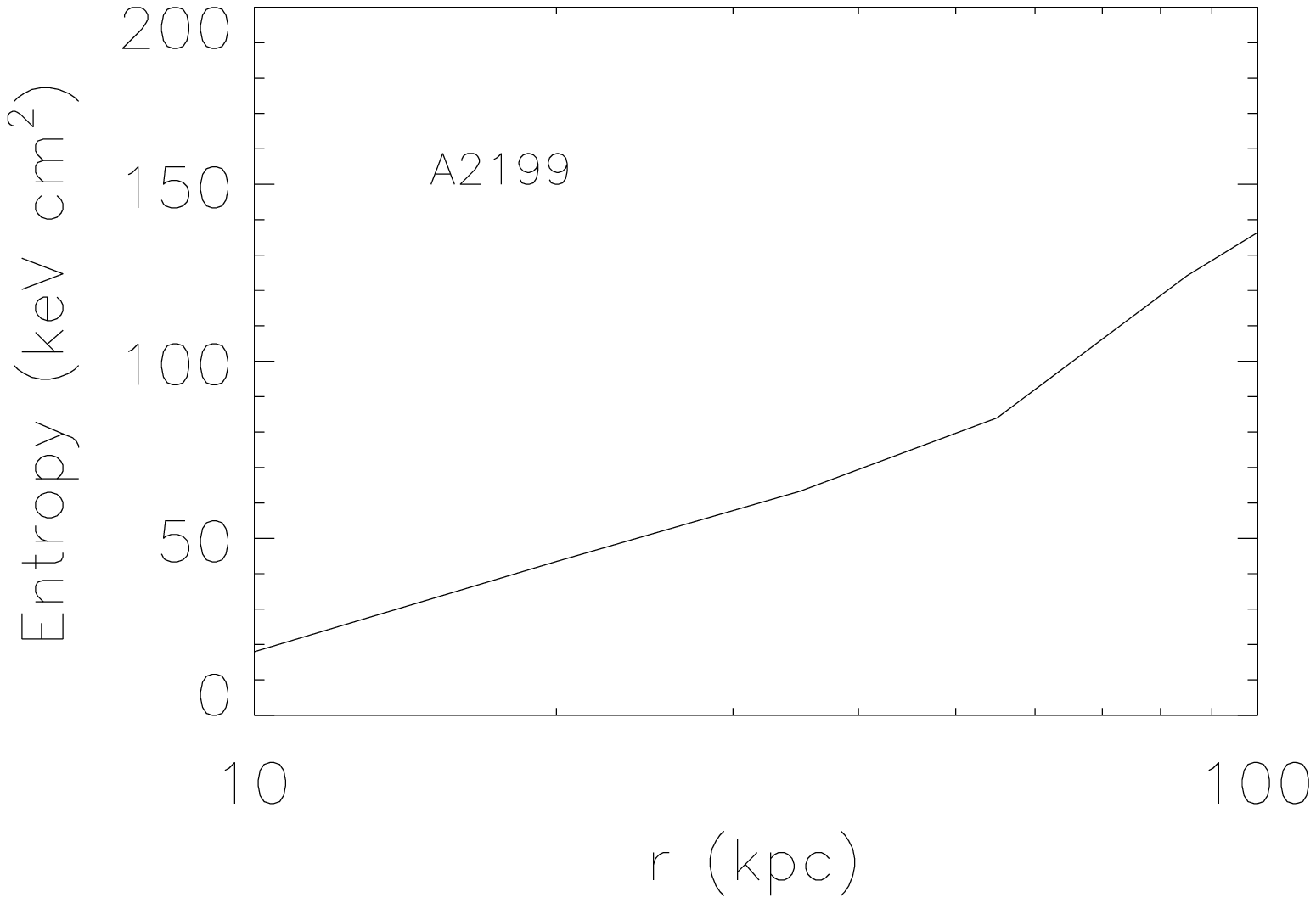,height=5.5cm,width=9.cm,angle=0.0}
 \epsfig{file=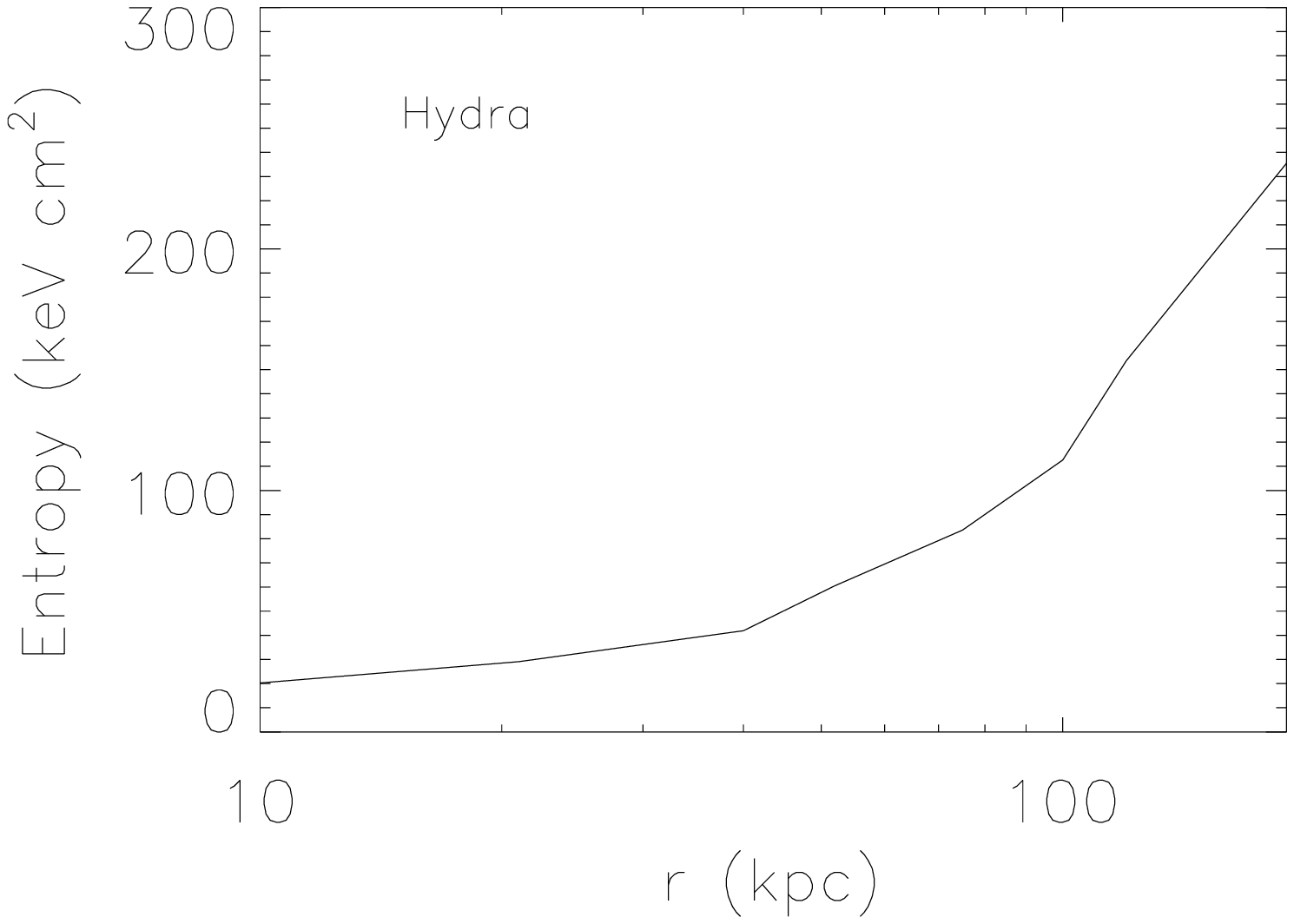,height=5.5cm,width=9.cm,angle=0.0}
 }
\end{center}
 \caption{\footnotesize{The radial profile of the IC gas entropy $S$ of Perseus (upper panel),
 A2199 (mid panel) and Hydra (lower panel) as predicted in the WR model. The curves reproduce
 the observed entropy profiles of the three clusters as derived by Churazov et al. (2003) for Perseus,
 by Caffi et al. (2004) for A2199 and by David et al. (2001) for Hydra. }}
 \label{fig.entropy}
\end{figure}

The WR heating in the cluster core can be also responsible for the flattening of the IC
gas entropy profile as observed in cool-core clusters. We have compared the entropy
profiles predicted in our model for Perseus, A2199 and Hydra to the available data for
these clusters (see e.g., Churazov et al. 2003, Caffi et al. 2004, David et al. 2001,
respectively) and we found a very good agreement (see Fig.\ref{fig.entropy}). This is
expected, indeed, because our model is able to reproduce the cluster temperature profile
given the observed density profile of the cluster and therefore their combination $S =
kT_{inner}/n_e^{2/3}$ which is referred to as the IC gas entropy. This analysis bring
further robustness to the ability of our model to explain the structure of clusters'
cores in terms of a balance between WR heating and IC gas cooling.

\section{Gamma-ray emission of cool core clusters}
 \label{sec.gamma}

The distribution of WRs that heat the IC gas to recover the observed temperature profiles
in their inner regions inevitably produces a gamma-ray emission due to various emission
mechanisms: i) $pp \to \pi^0 \to \gamma + \gamma$; ii) inverse Compton scattering (ICS)
due to secondary electrons produced by $pp \to \pi^{\pm} \to e^{\pm}$; iii) non-thermal
bremsstrahlung emission of the same secondary electrons.

From the WR spectra discussed in this paper (see Sect.4), we calculated the gamma-ray
emission produced by the decay of the $\pi^0 \to \gamma \gamma$ produced in the
interactions between relativistic protons and the thermal gas target protons (see, e.g.,
Blasi \& Colafrancesco 1999; Marchegiani et al. 2007 for the details of the
calculations).
The contribution of secondary electrons to the gamma-ray emission via ICS of CMB photons
and non-thermal bremsstrahlung is negligible w.r.t. the $\pi^0 \to \gamma \gamma$ flux
(see also Marchegiani et al. 2007) and, therefore, we do not consider them in the
following discussion.

Note that, at variance with previous works that considered the non-thermal (radio, HXR
and gamma-ray) emission of CRs in clusters' atmospheres (e.g., Colafrancesco \& Blasi
1998, Blasi \& Colafrancesco 1999, Miniati et al. 2001a, 2001b, Pfrommer \& Ensslin 2004,
Marchegiani, Perola \& Colafrancesco 2007, Pfrommer et al. 2007), the density of the WRs
is not a free parameter in our model, because it is fixed by the procedure presented in
Sect.2 used to fit the temperature profile of the cluster. Therefore, the cluster's
gamma-ray emission is univocally determined by this condition and it is strongly related
to the density and temperature structure of the cluster's cool core.

\begin{figure}[ht]
\begin{center}
 \epsfig{file=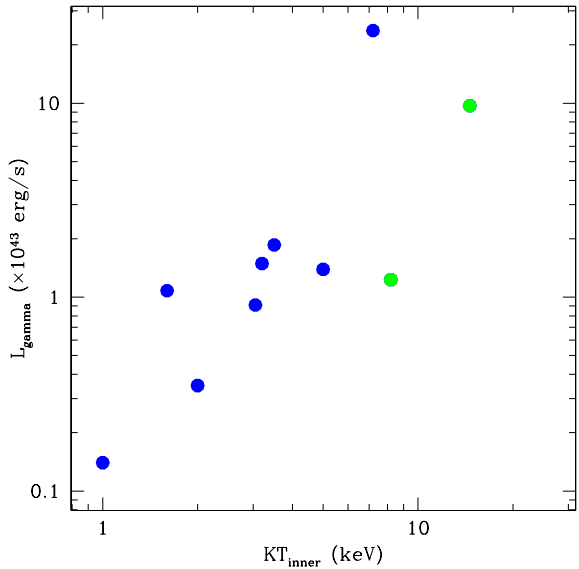,height=8.cm,width=8.cm,angle=0.0}
\end{center}
 \caption{\footnotesize{The correlation of the gamma-ray luminosity $L_{gamma}$ with the
 inner temperature of the cool core $T_{inner}$.
 Blue dots refer to cool-core clusters and green dots to non cool-core clusters.}}
 \label{fig.lg_t}
\end{figure}

The total gamma-ray flux and luminosity consistent with the cool-core temperature profile
of the considered clusters are reported in Table \ref{tab.2}.
The cluster gamma-ray luminosity predicted in our model shows an overall increase with
the value of $T_{inner}$ even though with some scatter.
The total cluster gamma-ray luminosity due to $pp \to \pi^0 \to \gamma \gamma$ decay
scales with the density of the WRs and of the IC gas according to  $L_{\gamma} \propto
n_{WR}(r) n_p(r) R^3$, and it is integrated out to the radius at which we see target
thermal protons from the clusters' X-ray images. The condition imposed by the WR heating
model to fit the inner temperature profile yields a scaling $L_{\gamma} \propto \sqrt{k
T_{inner}} (a/b) n^2_{e0} R^3$ which can be rewritten as $L_{\gamma} \propto \sqrt{k
T_{inner}} (a/b) L_X/(kT)^{1/2}$ and yields a scaling $L_{\gamma} \propto (k
T_{inner})^{\beta} (a/b)$ by using the observed $L_X \sim T^{\beta}$ relation (see Arnaud
2005 for a review) and a value $k T_{inner} \approx (1/3 - 1/2) kT$. The scaling of
$L_{\gamma}$ with $(k T_{inner})^{1.3}$ that derives form the previous considerations
reproduces quite well the overall increase of the gamma-ray luminosity of the ten
clusters in Tab.1 as a function of their inner core temperature $k T_{inner}$.\\
However, a scatter in the $L_{\gamma}-T_{inner}$ relation shown in Fig.\ref{fig.lg_t} is
indeed expected because the cluster gamma-ray luminosity depends on the different values
of $\alpha$ and on the different values of the integration volume $\propto R^3$ chosen
for each specific cluster. The choice of different integration volumes for the considered
clusters are motivated by the different extensions of the target protons in the thermal
gas to produce effective $pp \to \pi^0 \to \gamma \gamma$ collisions, as indicated by the
available X-ray data for each cluster (see Tab.1). A systematic increase of $L_{\gamma}$
with $T_{inner}$ is however shown by the data and can be understood by the previous
scaling-law considerations.

\begin{figure}[ht]
\begin{center}
 \epsfig{file=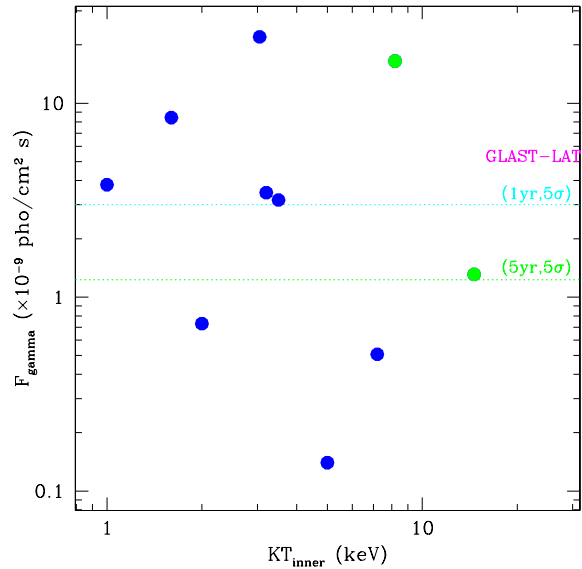,height=8.cm,width=8.cm,angle=0.0}
\end{center}
 \caption{\footnotesize{The distribution of the gamma-ray flux $F_{gamma}(E>0.1 \,\,\rm{GeV})$
 with the inner temperature of the cool core $T_{inner}$.
 Blue dots refer to cool-core clusters and green dots to non cool-core clusters.
 We show the GLAST-LAT sensitivity for a $5 \sigma$ detection after 1 yr (cyan line)
 and after 5 yrs (green line) (from Ritz 2007, on behalf of the GLAST collaboration
 see http://www-glast.slac.stanford.edu/software/IS/glast$_-$lat$_-$performance.htm).}}
 \label{fig.fg_t}
\end{figure}

We have also calculated the gamma-ray flux expected from each cluster in Table
\ref{tab.2} and we compared the results with the GLAST-LAT instrument sensitivity (see
Fig. \ref{fig.fg_t}). While all the gamma-ray fluxes are lower than the available EGRET
limits (Reimer et al. 2003), it is evident that 6 clusters with $F_{\gamma} \geq 3\times
10^{-9}$ photons cm$^{-2}$ s$^{-1}$ (A262, A2199, Perseus, Hydra, A1795 and Coma), and 7
clusters with $F_{\gamma} \geq 1.2 \cdot 10^{-9}$ photons cm$^{-2}$ s$^{-1}$ (the
previous clusters plus A2163) could be detected by the GLAST-LAT experiment in 1 year and
5 years at the $5\sigma$ confidence level, respectively.\\
The possible gamma-ray detection of these clusters will allow us to probe in details our
model for the WR heating of the cluster cores and consequently set constraints on the
parameters $n_{WR,0}$ and $\alpha$ of the WR heating model.

We stress that the gamma-ray emission evaluated here for cool-core clusters refer to the
truly diffuse component arising from $p p \to \pi^0 \to \gamma \gamma$ decay. We do not
consider here the possible contribution due to the gamma-ray emission from the central
engine of the radio galaxies and /or from their radio lobes. We will address more
completely this issue in a forthcoming paper.

Gamma-ray emission from WRs is able to probe directly the hadronic component of their
interaction with the IC gas. Nonetheless, the leptonic component of the WRs -- IC gas
interaction can be also efficiently probed by studying the secondary e$^{\pm}$ e.m.
emission: this is spread over a wide range of frequency, from radio (where the
synchrotron emission dominates) to the UV, X-ray and HXRs (where the ICS emission
dominates). We will discuss in the following these components separately.

\section{Diffuse radio emission in cool core clusters}
 \label{sec.radio}

Diffuse radio emission is also inevitably produced by synchrotron emission of the
secondary electrons produced by $pp \to \pi^{\pm} \to e^{\pm}$ interacting with the
intra-cluster magnetic field and it is therefore strongly related to the WR gamma-ray
emission and to the temperature profile of the cluster core. From the WR spectrum and
spatial distribution given in eq.(\ref{wr_spectrum}), we calculated the radio emission
produced by the secondary electrons (we follow the procedure described in Marchegiani et
al. 2007), by assuming a radial profile $B(r)$ for the magnetic field.

We apply this procedure to the clusters in Tab. \ref{tab.2} where a mini-halo is
observed, i.e. A2390, Perseus, \rxj\ and A2199 (in this last case an upper limit is only
available, see Kempner \& Sarazin 2000).

In the cluster A2390, we found that the intensity of the radio emission is
$2\times10^{-2}$ and $2$ mJy for values of central magnetic field in the range $1.3$ --
$13$ $\mu$G (which are equal to $B_{eq}$ and $10B_{eq}$, respectively, where $B_{eq}$ is
the value inferred by equipartion between magnetic field and electrons energy, Govoni \&
Feretti 2004), and for a radial shape of the magnetic field  $B(r) \propto n_e(r)$,
respectively. The radio flux associated to the A2390 mini-halo has been estimated to be
$63\pm 3$ mJy (Bacchi et al. 2003). However, the radio source at the center of A2390 is a
complex combination of the radio emission from the powerful FR-II flat-spectrum cD radio
galaxy and of the mini halo (Augusto, Edge and Chandler 2006). The substantial amount of
polarization detected around the cD radio galaxy and in the mini halo region indicates
that the radio flux attributed to the mini halo region could be substantially affected by
the radio galaxy lobes, therefore pointing to a quite lower flux of the truly diffuse
halo.

\begin{figure}[ht]
\begin{center}
\hspace{-1cm}
 \epsfig{file=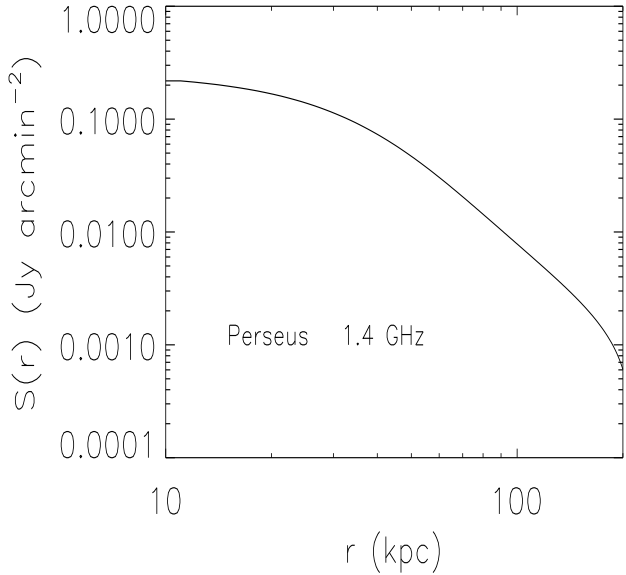,height=9.cm,width=9.cm,angle=0.0}
\end{center}
 \caption{\footnotesize{Radio brightness at 1.4 GHz emitted by secondary
electrons in Perseus cluster for $B_0=10$ $\mu$G and $B(r) \propto n_e^{0.5}$(r). The
density and the radial shape of the WRs  are those needed to stop the cooling flow at the
observed temperature level.}}
 \label{brilrad.perseus}
\end{figure}

For the Perseus cluster, we show in Fig. \ref{brilrad.perseus} the radio brightness at
1.4 GHz calculated for a magnetic field with central intensity of 10 $\mu$G and a radial
profile proportional to $n_e^{0.5}(r)$. These assumptions are comparable with those of
Pfrommer \& En\ss lin (2004), which have performed a calculation of the radio emission of
the secondary electrons in the Perseus cluster by fitting the CR density to obtain the
radio intensity. We found that, for the same magnetic field they used [i.e. $B_0 = 10$
$\mu$G and $B(r) \propto n^{0.5}_e(r)$], the WR density needed to quench the CF to the
observed temperature level reproduces very well the radio halo brightness (see Fig.
\ref{brilrad.perseus}).

\begin{figure}[ht]
\begin{center}
\hspace{-1cm}
 \epsfig{file=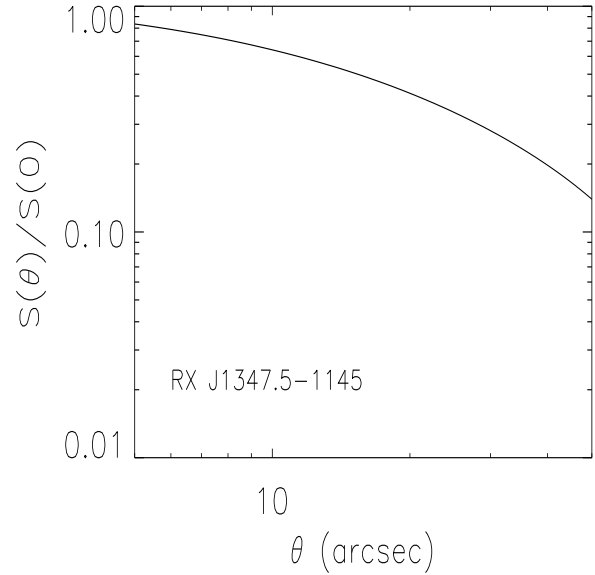,height=9.cm,width=9.cm,angle=0.0}
\end{center}
 \caption{\footnotesize{Normalized radio brightness emitted by secondary
electrons in the cluster \rxj\ evaluated at 1.4 GHz for $B_0=1$ $\mu$G and $B(r) \propto
n_e^{-0.5}(r)$. The density and the radial shape of the WRs are those needed to stop the
cooling flow at the observed level.}}
 \label{brilrad.rxj}
\end{figure}

For the cluster \rxj\ we found that the shape of radio emission (see Gitti et al. 2007)
can not be reproduced by secondary electrons if a monotonical radial decrease of the
magnetic field is assumed. Instead, if we assume $B_0=1$ $\mu$G and a profile of $B(r)
\propto n_e^{-0.5}(r)$, we obtain the brightness shape shown in Fig. \ref{brilrad.rxj}.
This profile is very similar to that of Gitti et al. (2007), and the total flux
calculated is 33 mJy, comparable to the measured one of $\sim 25$ mJy. It will be easy to
recover the observed radio halo flux with a central magnetic field $\simlt 1$ $\mu$G.

In A2199, where we use a radial profile $B(r) \propto n_e(r)$, we obtain a total flux of
100 mJy for $B_0=5$ $\mu$G and 372 mJy for $B_0=10$ $\mu$G. Since the upper limit is 168
mJy (see Kempner \& Sarazin 2000), and these values of the central magnetic field are
comparable with those obtained by FR measurements in cool--core clusters, we conclude
that the available radio data are compatible with our model, if the central magnetic
field is not very strong, i.e. for values of $B_0\simlt 7$ $\mu$G.

For the two non cool-core clusters with radio halos considered in our analysis, we
recover the results already presented in previous analyses (Marchegiani, Perola \&
Colafrancesco 2007). Specifically, for Coma we found $\alpha = 1$, i.e. the condition
$P_{WR} \propto P_{th}$, and the same ratio $P_{WR}/P_{th} \approx 0.3$ (indicated also
by the simulations of Ryu et al. 2003) for which the radio halo surface brightness can be
recovered with a central magnetic field of 1.2 $\mu$G (Marchegiani et al. 2007).\\
For A2163 we again found $\alpha = 1$, i.e. the condition $P_{WR} \propto P_{th}$, and a
ratio $P_{WR}/P_{th} \approx 0.2$ for which the radio halo surface brightness can be
recovered with a central magnetic field of 4.5 $\mu$G, a factor $\sim 2$  higher than
that found by Marchegiani et al. (2007).

We note that in clusters where no diffuse radio emission has been detected, the magnetic
field required by our WR model should be lower than the value required to make the radio
emission from secondary electrons detectable. In fact, the synchrotron emission depends
non-linearly on the magnetic field intensity and fluctuations (see discussion in
Colafrancesco, Marchegiani \& Perola 2005) and therefore, little or moderate variations
of its amplitude can decrease the radio flux by even a large amount. It is important to
remark that future, deeper radio observation of galaxy clusters, as those which can be
obtained with LOFAR and/or SKA, can provide further information on the intensity of the
radio emission, and on the non-thermal content of galaxy cluster (see discussion in
Marchegiani et al. 2007).

We note also that the WRs model can provide a consistent picture for the formation of
morphologically different radio sources in galaxy clusters: in fact, the diffusion
processes in different clusters can produce different WRs radial profiles, then
mini-halos or giant radio halos can be produced. In addition, the action of localized
external compression (i.e. by action of merging) can enhance the CRs density and the
magnetic field intensity in external regions of the clusters, by producing the radio
relics.

\section{Diffuse Hard X-Ray emission in cool core clusters}
 \label{sec.hxr}

The secondary electrons from WR interactions with the IC gas protons also produce
radiation in the Hard X-Ray band, by Inverse Compton Scattering (ICS) of the Cosmic
Microwave Background (CMB) Radiation photons.

The equilibrium spectrum of the secondary electrons depends on the density of the WRs
(which is determined in our approach) and on the energy loss rates of the electrons (see,
e.g., Marchegiani, Perola \& Colafrancesco 2007). While the ICS losses are known and are
uniform all along the clusters volume, the synchrotron losses depend on $B^2(r)$, and
therefore they are not known, in principle. However, for the clusters which host a
diffuse radio emission, information on the magnetic field $B(r)$ can be inferred from the
observed radio-halo features (see Sect.6 above), while for the other clusters it is
necessary to assume a magnetic field structure.

Therefore, to predict the HXR emission of clusters listed in Tab. 1, we adopt two
different strategies: \textit{i)} for the radio emitting clusters we calculate the HXR
emission by using a magnetic field compatible with the observed radio emission (see
Sect.\ref{sec.radio}); \textit{ii)} for all the clusters we calculate the HXR emission by
assuming a low magnetic field as generally indicated by the ICS interpretation of the HXR
emission observed in nearby clusters.  In this last case, the synchrotron losses are
everywhere weaker that the ICS ones, and the density of the secondary electrons -- and
consequently the HXR emission -- have to be considered the maximum ones which is
compatible with the cool--core heating by the WRs.
The HXR fluxes reported in Tab.2 refer to this last evaluation strategy.

\begin{figure}[ht]
\begin{center}
\hspace{-1cm}
 \epsfig{file=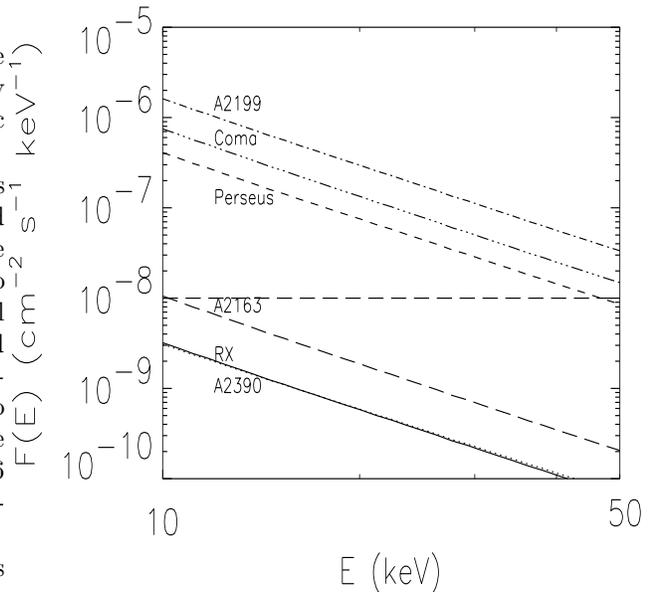,height=9.cm,width=9.cm,angle=0.0}
\end{center}
 \caption{\footnotesize{ICS emission in the HXR band of the secondary
electrons for the radio emitting clusters. RX stands for \rxj. The horizontal,
long-dashed line represents the approximate sensivity of Simbol-X in the $10-50$ keV
range. The HXR fluxes have been computed under strategy \textit{i)} (see text for
details).}}
 \label{hxr_cf}
\end{figure}

In Figure \ref{hxr_cf} the ICS emission spectrum of the secondary electrons is shown for
the radio emitting clusters (A2199, Coma, Perseus, A2163, A2390, \rxj). These fluxes --
referring to the evaluation strategy \textit{i)} -- can be compared with the expected
sensitivity in the HXR band of the next coming mission Simbol-X (Ferrando et al. 2005),
which is approximately $\sim10^{-8}$ photons cm$^{-2}$ s$^{-1}$ keV$^{-1}$ in the energy
band $10-50$ keV. From this Figure we can conclude that the clusters A2199, Coma and
Perseus should be detectable by Simbol-X, if the magnetic fields in these clusters are
those inferred by assuming that the radio emissions are produced by secondary electrons.
The HXR emission expected in the 10--50 keV band should be clearly disentangled from the
thermal one in this energy band. On the contrary, the clusters A2163, \rxj\ and A2390
should not be detectable by Simbol-X in this evaluation strategy (for the case of A2390
we have used a magnetic field value $\simeq 10 B_{eq}$, see discussion in Sect.6).

\begin{figure}[ht]
\begin{center}
\hspace{-1cm}
\vbox{
 \epsfig{file=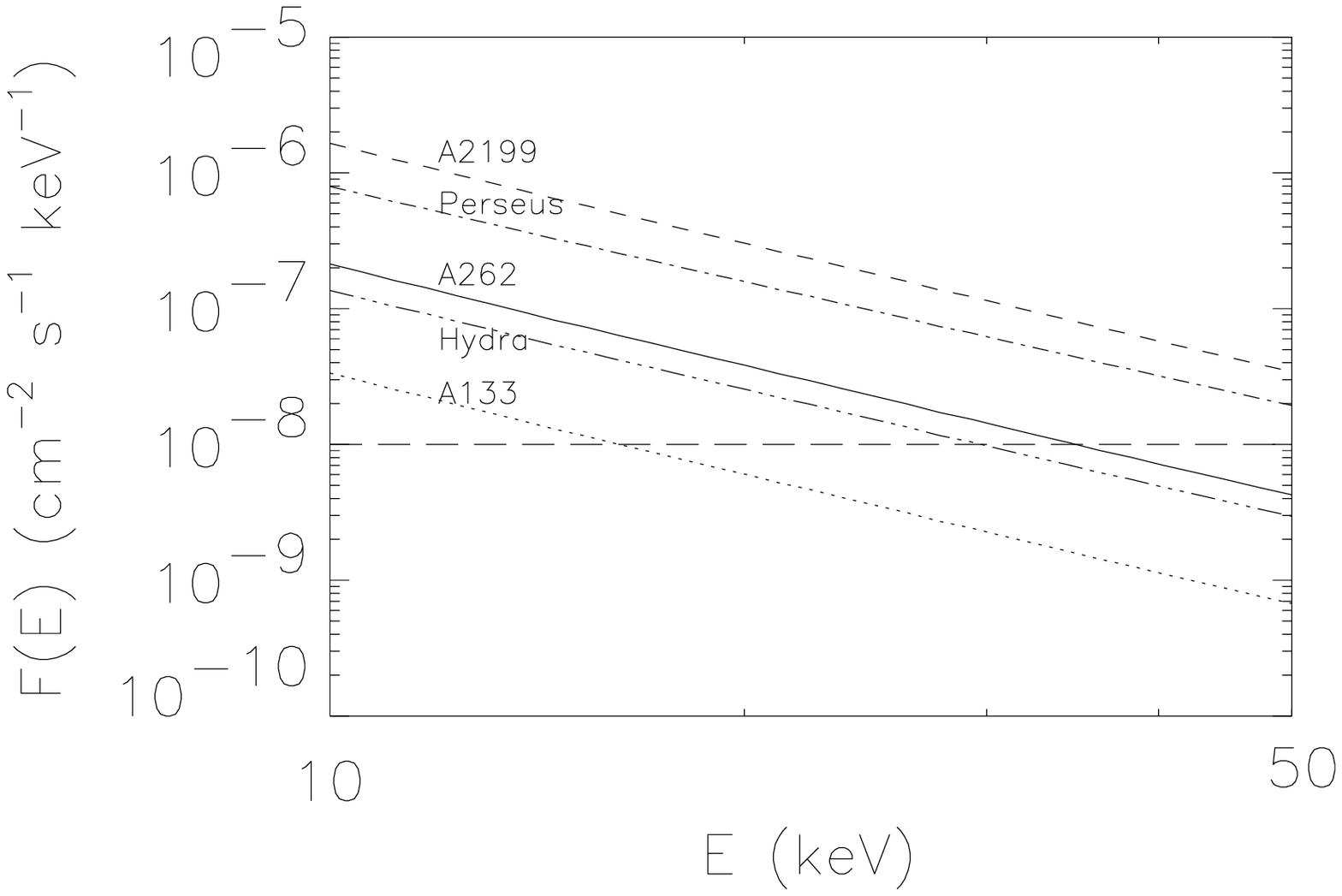,height=9.cm,width=9.cm,angle=0.0}
 \epsfig{file=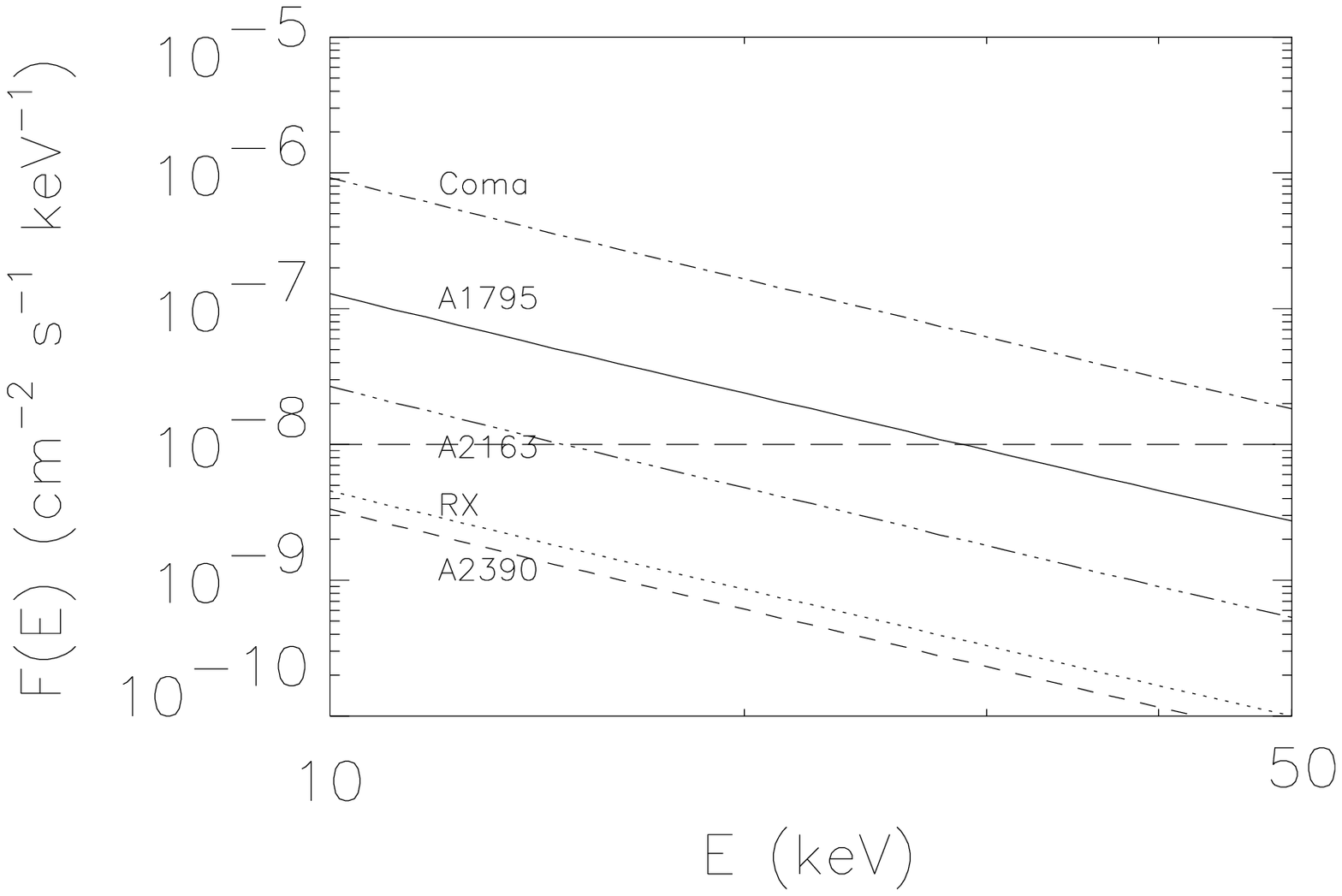,height=9.cm,width=9.cm,angle=0.0}
 }
\end{center}
 \caption{\footnotesize{ICS emission in the HXR band of the secondary
electrons for all the clusters considered. For clarity, the emissions are reported in two
different panels. RX stands for \rxj. The horizontal, long-dashed line represents the
approximate sensitivity of Simbol-X in the $10-50$ keV range. The HXR fluxes have been
computed here under strategy \textit{ii)} (see text for details).}}
 \label{hxr_max_cf}
\end{figure}

In Fig.\ref{hxr_max_cf} (both panels), we show the HXR ICS emission of the secondary
electrons in the WR model for all the clusters we considered (see Table \ref{tab.1}), by
assuming a value of the magnetic field $B_0=0.1$ $\mu$G, for which energy losses are
completely dominated by ICS, i.e. our evaluation strategy \textit{ii)}. Note that these
levels of HXR emission should be considered as upper limits. From this figure, we can
conclude that only \rxj\ and A2390 should be not detectable by Simbol-X; a difficult
detection is possible for A133 and A2163, while the other six clusters should be
definitely detectable by Simbol-X, if the extreme conditions on the magnetic field that
we have assumed hold.

The HXR fluxes predicted in the WR model are also much lower than the 20-80 keV fluxes of
the nearby clusters detected with BeppoSAX (see Nevalainen et al. 2004). Specifically, we
obtain for Coma a flux $F_{20-80 keV}= 1.6 \cdot 10^{-13}$ erg cm$^{-2}$ s$^{-1}$ (vs. an
observed flux of $F_{20-80 keV}= (1.1 \pm 0.5) \cdot 10^{-11}$ erg cm$^{-2}$ s$^{-1}$),
for A2199 a flux $F_{20-80 keV}= 2.1 \cdot 10^{-13}$ erg cm$^{-2}$ s$^{-1}$ (vs. an
observed flux of $F_{20-80 keV}= (8.4 \pm 3.9) \cdot 10^{-12}$ erg cm$^{-2}$ s$^{-1}$)
and for A2163 a flux $F_{20-80 keV}= 3.2 \cdot 10^{-15}$ erg cm$^{-2}$ s$^{-1}$ (vs. an
observed flux of $F_{20-80 keV}= (1.7 \pm 3.1) \cdot 10^{-12}$ erg cm$^{-2}$ s$^{-1}$).
The ICS HXR fluxes yielded by the WRs in cluster cores cannot reproduce the HXR fluxes
observed towards these clusters with BeppoSAX (see also Colafrancesco, Marchegiani \&
Perola 2005 for a general, phenomenological discussion on this point).

We also evaluated the non-thermal flux produced by ICS of the secondary electrons in the
2-10 keV energy band intergrated in  a $\sim 250$ arcsec region around the center of
Perseus to be compared with the thermal bremsstrahlung emission of the IC gas in the same
region of the cluster. The non-thermal ICS flux induced by WRs is a factor $\sim  10^3$
less than the thermal bremsstrahlung flux. This further confirms that ICS HXR fluxes
yielded by the WRs in cluster cores cannot reproduce the non-thermal flux observed
towards the centers of clusters.

\section{Discussion}
 \label{sec.discussion}

We have shown that a specific feature of the WR scenario for the heating of cool cores
in galaxy clusters is that it is inevitably associated to emission features that manifest
themselves over a wide range of frequencies, from radio to gamma-rays.\\
This fact provides a clear test to distinguish our model from other models of
cooling-flow quenching of mechanical and/or non-gravitational origin as well as from
other models for the origin of non-thermal phenomena in the inner regions of galaxy
clusters.

How the WR model presented here compares with these models?\\
Guo \& Oh (2007) proposed a model of AGN feedback heating, in which cosmic rays produced
by accretion triggered AGN activity heat the IC gas efficiently, with small dynamical
perturbations on the IC gas itself. This model considers that the ICM is heated by
cosmic-rays, which are injected into the ICM through diffusion or the shredding of the
bubbles by Rayleigh-Taylor or Kelvin-Helmholtz instabilities, with the inclusion of
thermal conduction.
It seems that the inclusion of CRs in this model solves the extreme fine tuning required
by the conduction--suppression factor when only thermal conduction operates, and provides
a more stable radial distribution of IC gas temperature that is, therefore, mainly driven
by the effect of CRs which set some low value of the mass accretion rate with little
additional effect of the thermal conduction (see their Fig.3 and their discussion).

The large impact of CRs as the main heating and regulating agent of the clusters'
cool-cores is quite similar to what is found in our model, because the conduction effects
are quite marginal.
However, our WR model requires quite substantial WR pressures at the cluster's center
(that becomes negligible in the outer regions) that are able to recover the non-thermal
pressure profile found in Perseus and help reproducing the mini radio halo surface
brightness of cool-core clusters (Perseus, \rxj, A2199) with reasonable values and radial
profiles of the cluster's magnetic field.
The low CR pressure ($P_{CR} \simlt 0.1 P_{th}$) and energy required by the model of Guo
\& Oh (2007) in A2199 can be explained as an effect of the lower energy cutoff ($E_{min}
\approx 706$ MeV) in the CR spectrum which is much higher than our choice ($E_{min} = m_p
c^2 \cdot [1+3.4 \cdot 10^{-5}(kT/keV)]$ (see Sect.3). We have quantitatively shown this
effect in the specific case of A2199 (see Fig.6) where we compare our WR energy density
distribution with that of the CR feedback heating model of Guo \& Oh (2007).
Such difference in the energy spectrum of CRs causes also a different level of predicted
gamma-ray fluxes for clusters with cool-cores, with our WR model providing higher fluxes
detectable with the GLAST-LAT experiment and with the next generation HXR experiments
like Simbol-X.

In particular, the Perseus cluster should be one of the best candidates to look for a HXR
ICS emission feature in the 10-50 keV range emerging from its inner core (see Figs.13 and
14). This prediction seems quite robust since the WR distribution responsible for such
ICS emission is the same that is able to reproduce the thermal gas temperature
distribution in its cool core and the mini radio halo surface brightness visible in the
same spatial region. This prediction is also strengthened by the ability of the WR model
to recover at the same time the thermal gas temperature radial distribution and the
non-thermal (WR) pressure radial profile observed in the Perseus core (see Sanders \&
Fabian 2007).\\
Other clusters with radio halos (e.g. Coma and the marginal detection of A2199) are also
predicted to be visible by Simbol-X once we normalize our model to their temperature
profile and radio-halo emission.

WR (CR) heating models are a viable alternative to mechanical heating models (e.g.
Br\"uggen \& Kaiser 2002; Ruszkowski et al. 2004; Vernaleo \& Reynolds 2006; Reynolds et
al. 2005, McCarthy et al. 2007) in which the ICM is heated by the $p dV$ mechanical work
of the expanding bubbles, viscous dissipation of emitted sound waves or mixing of the hot
bubble plasma with the ICM.
The detailed microphysics of how the mechanical heating processes take place has not been
hammered out in detail, leaving a good deal of uncertainty; a definitive explanation for
how energy is transported from the observed bubbles to the ICM in a distributed and
isotropic fashion is still outstanding.
The possible role of merging shocks to the heating of the IC gas in the clusters' cores
is still to be quantitatively determined (see the simulations of Ryu  et al. 2003).

It is clear, nonetheless, that also cosmic-ray heating models suffer from other
uncertainties: the details of how WRs adjust to a density profile similar to that of the
IC gas have to be addressed by specific studies that could bring further robustness to
the preliminary calculations presented here (see Appendix A and Sect.2).\\
It is possible to envisage various scenarios for the WR origin in terms of the activity
of AGNs with small duty cycles and/or a substantial fractions of AGNs in cluster cores
(see, e.g., Bird et al. 2007), or CR origin in non-standard scenarios like cannonball
models (see deR\'ujula 2004 for  review), but we do not want to address this specific
topic here.
However, it seems that the WR model is somewhat less uncertain that the CR feedback
heating, for which large uncertainties stand on how cosmic rays leak from the bubbles,
and/or the rate at which bubbles are disrupted. It is nonetheless important to stress
that elucidating the details of bubble disruption/cosmic-ray diffusion would be very
useful in determining whether cosmic-rays or mechanical processses provide a more
efficient means of transporting heat from the bubble to the ICM.

All of these issues may assume a great relevance in the light of the experimental
possibility to study CR physics in large-scale structures in a multi-frequency approach:
the next coming gamma-ray observations of several nearby clusters with GLAST, the future
HXR detection of the ICS emission from the same clusters with Simbol-X and the possible
detection of non-thermal SZ effect signatures (see, e.g. Colafrancesco, Marchegiani \&
Palladino 2003, Colafrancesco 2007 for a review) in the spectra of these clusters with
SPT will clearly help to disentangle not only the presence of cosmic rays in the cluster
atmospheres, but also the details of various WR models for the heating of cluster
cores.\\
In fact, models of mechanical heating do not show any non-thermal emission feature while
the WR model predict a substantial level of non-thermal emission at various frequency
(radio, microwave, HXR and gamma) and the CR feedback heating model of Guo \& Oh (2007)
predicts only marginally visible non-thermal phenomena in cool-core clusters.\\
The next generation radio, HXR and gamma-ray experiments will certainly disentangle these
two classes of models and will determine the amount of WRs which are present in galaxy
clusters, provided that a good separation of the diffuse and point-like non-thermal
emission will be achievable.


\section{Conclusions}
 \label{sec.conclusions}

The WR heating model that we presented in this paper is able to reproduce the
temperature, the pressure and the entropy structure of both cool-core and non cool-core
clusters and it provides results that are consistent with all the available evidence of
non-thermal phenomena emerging from the cores of clusters, e.g. diffuse radio emission in
the form of mini halos and diffuse radio halos, HXR emission limits obtained with
BeppoSAX, Chandra and XMM, and gamma-ray limits obtained with EGRET.

This model provides a theoretical description of the physics of cool cores that is
directly related to the presence of several observable consequences of the presence of
WRs in clusters' atmospheres. The main conclusions of our work are:
\begin{itemize}
\item the presence of WRs in cluster atmosphere provides a simple solution for the
temperature, pressure and entropy profiles of the thermal IC plasma and of the
non-thermal plasma observed in the inner regions of several clusters studied in this
paper.
\item The WR distribution produces a substantial emission of gamma-rays by their
hadronic $pp \to \pi^0 \to \gamma \gamma$ interations with the IC gas. A large fraction
(6 clusters out of 10 considered in this study) have gamma-ray fluxes at $E > 100$ MeV
detectable with the GLAST-LAT experiment. The gamma-ray luminosity of the cool-core and
non-cool core clusters correlates with the cluster's inner temperature, and thus provides
a specific scaling behaviour of our WR model that can be verified with the next coming
GLAST observations.
\item The WRs distribution can reproduce the radio emission of all the cool-core
clusters with mini-halos that we studied (A2199, Perseus, \rxj) except for A2390 (whose
mini radio-halo flux could be affected by residual radio emission from the lobes of the
central cD radio galaxy).
\item The WRs distribution produces ICS emission with fluxes well below the HXR limits of
A2199, Coma and A2163 provided by BeppoSAX and well below the not-thermal emission
detected by Chandra and XMM in the central regions of Perseus. This means that ICS
emission from WRs is not the explanation of the emission of these clusters in excess with
respect to their thermal bremsstrahlung radiation.
\item The distribution of WRs in clusters directly relates the thermal IC gas
temperature, pressure and entropy distribution in cool-core clusters (observable in the
X-ray energy band) with non-thermal diffuse emission features observable at radio
(synchrotron), HXRs (ICS), gamma-rays (mainly from $\pi^0 \to \gamma \gamma$ decay) and
microwaves (by Sunyaev --Zel'dovich effect). We predict that the expected levels of these
non-thermal emission features in the WR model will be testable with the next-coming
experiments in the HXR band (Simbol-X), in the gamma-rays (GLAST-LAT) and in the
microwaves (SPT). These experiments can determine the amount of WRs which are present in
galaxy clusters.
\item These observable predictions makes the WR model entirely testable by using a
multi-frequency observational strategy that links, therefore, thermal and non-thermal
phenomena in clusters cores.
\item The specific theoretical and observational features of the WR model render it quite
different from other models so far proposed for the heating of cool-cores in galaxy
clusters. Such peculiar differences make it possible to prove or disprove the WR model
and, in general, the properties of any model that has been proposed as an explanation of
the cooling-flow problems.
\end{itemize}

\begin{acknowledgements}
We thank the Referee of this paper for various suggestions that helped in improving the
presentation of the manuscript.

\end{acknowledgements}

\appendix

\section{CR proton diffusion in cluster atmospheres}

We calculate the effect of the diffusion of CR protons (we remind the reader that we use
the same notation for CRs and WRs) in the cluster atmosphere according to the following
equation:
 \be
 {\partial N \over \partial t} - \nabla (D\nabla N)- {\partial
 (b_p N) \over \partial E} = Q_p,
 \label{eq.pdiff}
 \ee
where $N(E_p,r)$ is the proton CR density, $D(E_p,r)$ is the diffusion coefficient,
$b_p(E_p,r)$ is the proton energy loss term and $Q_p(E_p,r)$ is the source term.

Protons we are interested in here do not loose appreciably their energy
on time scales of the order of the cluster's lifetime
so that we can
neglect the term $\partial (b_p N) / \partial E $ in eq.(\ref{eq.pdiff}). We are
interested in a quasi-stationary solution (${\partial N / \partial t} =0$),
so that the diffusion equation simplifies to
 \be
  \nabla (D\nabla N) = -Q_p,
 \label{eq.pdiffstationary}
 \ee
which, in spherical simmetry, can be written as:
\be
\frac{1}{r^2}\left[ \frac{\partial}{\partial r}\left(r^2 D \frac{\partial N}{\partial
r}\right)\right]=-Q_p. \label{eq.diffu.3} \ee The solution of this equation writes as:
\be
N(r)=\int \left[ \frac{-\int dr r^2 Q_p}{r^2 D} \right] dr. \ee
 We search for a radial
shape of the source term $Q_p(E_p,r)$ which, by the diffusion effect, gives a radial
shape of the proton equilibrium distribution which is proportional to $[g(r)]^\alpha$
(this is the radial dependence of the WRs to be inserted in eq.(\ref{Tevol})), where
$g(r)$ is the radial profile of the gas. Therefore, we write:
\be
\int  \frac{I_1(r)}{r^2 D} dr \equiv N_0 [g(r)]^\alpha, \ee where
\be
I_1(r)=-\int_0^r dr r^2 Q_p(r).
\ee
This is a general solution of eq.(\ref{eq.pdiff}) under the assumed hypothesis.
Under the specific assumption that the diffusion coefficient
is constant with the radius, we can write
\be
\frac{1}{D_0} \int_0^r d r' \frac{I_1(r')}{r'^2}= N_0 [g(r)]^\alpha. \ee By deriving both
members with respect to $r$, we obtain:
\be
\frac{1}{D_0} \frac{I_1(r)}{r^2}= N_0 \alpha [g(r)]^{\alpha-1} \frac{d}{dr}g(r), \ee and
then
\be
I_1(r)= N_0 D_0 \alpha h(r), \ee where
\be
h(r)=r^2 [g(r)]^{\alpha-1} \frac{d}{dr}g(r). \label{eq.h} \ee
 By deriving with respect to $r$, we obtain
\be
r^2 Q_p(r)=- N_0 D_0 \alpha \frac{d}{dr}h(r) \ee
 and then we obtain the source term that satisfies eq.(\ref{eq.diffu.3}):
 \be
Q_p(r)=-N_0 D_0 \alpha \frac{1}{r^2}\frac{d}{dr}h(r). \label{eq.qp}
 \ee
This equation gives the source term $Q_p(E_p,r)$ which produces a radial distribution of
protons proportional to $[g(r)]^\alpha$, for a constant diffusion coefficient $D(r)\equiv
D_0$. In conclusion, we have shown that it is possible, by knowing the radial shape of
the thermal gas in a cluster, to know the radial shape that the WR source term should
have in order to recover the required radial distribution of the warming rays.

\begin{figure}[ht]
\begin{center}
\hspace{-1cm}
 \epsfig{file=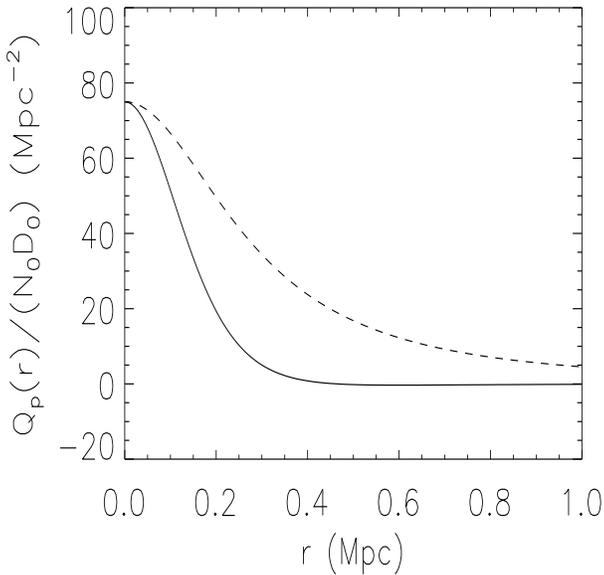,height=9cm,width=9cm,angle=0.0}
\end{center}
 \caption{\footnotesize{Source term [eq.(\ref{eq.qp})] for the
 case of a cluster with the thermal profile as Coma ($r_c=0.3$ Mpc
 and $q_{th}=1.125$) and for $\alpha=1$. With the dashed line, it
 is drawn the thermal profile of the cluster (which, in this case,
 is also the resulting proton radial distribution), with an arbitrally
chosen normalization for a better comparison. }}
 \label{figura.sorgente}
\end{figure}

In the following, we consider a solution of eq.(\ref{eq.diffu.3}) for a specific case of
$g(r)$, and we find the corresponding shape of the source term $Q_p(r)$. We consider only
the radial shape of the proton distribution, and we use a free normalization.\\
We consider a thermal IC gas profile of the form
 \be
g(r)=\left[1+\left(\frac{r}{r_c}\right)^2\right]^{-q_{th}}.
 \ee
Equations (\ref{eq.qp}) and (\ref{eq.h}) provide the shape of the source term $Q_p(r)$.
In Fig.\ref{figura.sorgente}, we consider the case of a cluster with $r_c=0.3$ Mpc and
$q_{th}=1.125$ (corresponding to the case of Coma). As we can see from this figure, the
effect of the proton diffusion is to provide a broad equilibrium distribution of CRs from
a narrow distribution of the proton source term.

\begin{figure}[ht]
\begin{center}
\hspace{-1cm}
 \epsfig{file=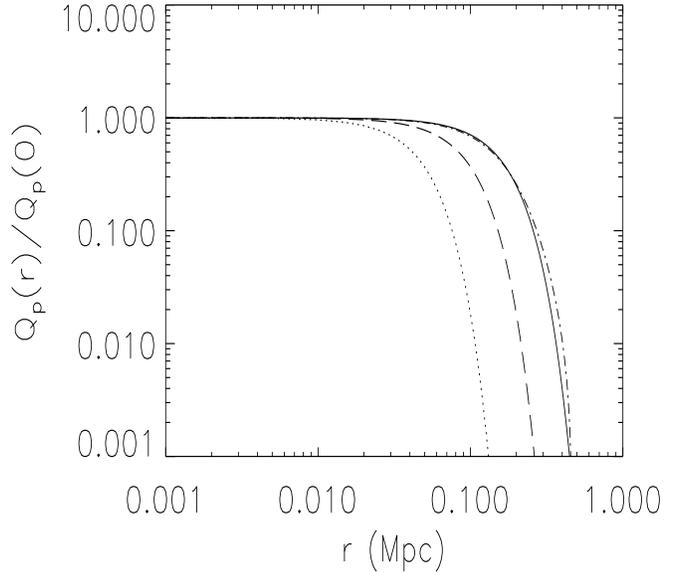,height=9cm,width=9cm,angle=0.0}
\end{center}
 \caption{\footnotesize{Source term (dot-dashed line) as in Fig.\ref{figura.sorgente}
 compared with a Gaussian shape as in eq.(\ref{eq.gaussiana}) with
 $r_s=0.17$ Mpc (solid line), $r_s=0.10$ Mpc (dashed line),
 and $r_s=0.05$ Mpc (dotted line).
}}
 \label{figura.sorgente2}
\end{figure}

A Gaussian distribution
 \be
 Q_p(r) \propto \exp \left[-\left(\frac{r}{r_s} \right)^2 \right],
 \label{eq.gaussiana}
 \ee
with a small value of $r_s$ is able to reproduce the source term required to provide the
CR profile $[g(r)]^\alpha$ with $\alpha=1$. In Fig.\ref{figura.sorgente2}, we compare the
source term in Fig.\ref{figura.sorgente} with a Gaussian distribution as in
eq.(\ref{eq.gaussiana}) with $r_s=0.17$ Mpc and two other Gaussians with $r_s=0.10$ and
$r_s=0.05$ Mpc. Then this function, with properly chosen parameters (normalization and
$r_s$), can be a good approximation for the required source term.

\begin{figure}[ht]
\begin{center}
\hspace{-1cm}
 \epsfig{file=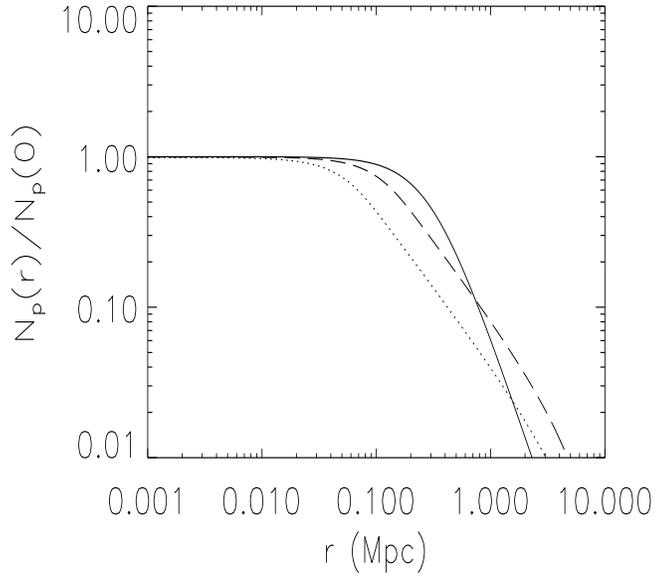,height=9cm,width=9cm,angle=0.0}
\end{center}
 \caption{\footnotesize{Equilibrium distribution of the protons
 produced by a Gaussian source term as in eq. (\ref{eq.gaussiana})
 with $r_s=0.17$ Mpc (solid line), $r_s=0.10$ Mpc (dashed line),
 and $r_s=0.05$ Mpc (dotted line).
}}
 \label{figura.sorgente3}
\end{figure}

In Fig. \ref{figura.sorgente3} we show the equilibrium distribution of the CR protons
produced by several Gaussian source terms as in eq. (\ref{eq.gaussiana}) with $r_s=0.17$,
0.10 and 0.05 Mpc, and we find that the first function is very similar to the IC gas
radial distribution of a Coma-like cluster. Since this procedure is not an accurate fit,
but only a qualitative comparison for an arbitrary set of parameters, we can conclude
that a narrow CR proton source term, by effect of the spatial diffusion, can produce a
broad radial equilibrium distribution of the CR protons which is very close to the
thermal IC gas distribution.


\end{document}